\documentclass[a4paper,11pt]{article}
\usepackage{amsfonts}
\usepackage{amsmath}
\usepackage{amssymb}
\usepackage{graphicx}
\usepackage{array}
\usepackage{booktabs}
\usepackage{multirow}
\usepackage{makecell}
\usepackage{float}
\pdfoutput=1 

\usepackage{subcaption}
\usepackage{mymacros}
\usepackage{chngcntr}
\usepackage{verbatim}
\usepackage{amsthm}
\usepackage{graphics}
\usepackage{xcolor}
\usepackage{mathrsfs}
\usepackage{bbold}
\usepackage{cases}
\usepackage{epsfig}
\usepackage{epstopdf}
\usepackage{hyperref}
\usepackage{amsfonts}
\usepackage{hyperref}
\usepackage{jheppub} 

\usepackage[T1]{fontenc} 

\hypersetup{
	colorlinks=true,
	linkcolor=blue,
	filecolor=blue,
	urlcolor=blue,
	citecolor=blue,
}

\title{\boldmath Chaos bound for spinning particles in Kerr-Newman-AdS black holes}

\author{Deyou Chen \footnote{E-mail: deyouchen@hotmail.com },}
\author{Chuang Yang \footnote{E-mail: chuangyangyc@hotmail.com},}
\author{Kangqiao Liu \footnote{E-mail: kqliu@xhu.edu.cn}}

\affiliation{School of Science, Xihua University, \\No. 999 Jinzhou Road, Pidu District, Chengdu 610039, China.}

\abstract{In this paper, we investigate the chaos bound violation of spinning charged test particles in the Kerr--Newman--AdS spacetime by comparing the Lyapunov exponent of unstable circular orbits with the black hole surface gravity. We find that the violation is governed by the relative balance between the local orbital-instability and surface gravity scales, while spacetime rotation, charge, AdS curvature, and particle spin modify this balance in different ways. In rotating backgrounds, the violation depends strongly on the relative orientation between the particle total angular momentum and the black hole rotation, with anti-aligned configurations generally favoring violation. Particle spin further shifts the violation boundary through spin--curvature coupling, although its influence remains comparatively weak over the spin range considered. The cosmological constant has a background-dependent influence. These results show that the parameter dependence of chaos bound violation reflects the competition between the local orbital-instability and surface gravity scales.}

\begin{document} 
	\maketitle
	\flushbottom
	
\section{Introduction}\label{sec1}

Chaos is a fundamental feature of nonlinear dynamical systems, characterized by a strong sensitivity to initial conditions that is commonly quantified by the Lyapunov exponent (LE). For thermal quantum systems with a large number of degrees of freedom, Maldacena, Shenker, and Stanford conjectured a universal temperature-dependent upper bound (chaos bound) on the LE \cite{MSS},

\begin{eqnarray}
\lambda \leq \frac{2\pi T}{\hbar},
\label{eq1.1}
\end{eqnarray}

\noindent where $T$ is the temperature of the system and $k_{B}=1$ is assumed. The conjecture has been extensively studied using gravitational shock-wave analyses and within the framework of the AdS/CFT correspondence \cite{SS1,SS2,JMM}. Since its proposal, this pioneering work has garnered widespread attention and substantial support. A notable example is the Sachdev--Ye--Kitaev (SYK) model, which saturates the bound in the strongly coupled regime \cite{SYK1,SYK2,SYK3,SYK4}. Saturation is also obtained for holographic systems described by classical Einstein gravity under appropriate conditions \cite{SS3}.

For classical systems, Hashimoto and Tanahashi investigated the chaotic behavior of particle motion near the horizon of a spherically symmetric black hole (BH) \cite{HT}. They found that the corresponding LE calculated at the maximum of the unstable effective potential is independent of both the particle species and external forces acting on the particle. Instead, it is determined solely by the BH surface gravity,

\begin{eqnarray}
\lambda = \kappa.
\label{eq1.2}
\end{eqnarray}

\noindent Through the relation between the Hawking temperature and the surface gravity, this result provides a classical counterpart of the quantum chaos bound and motivates the inequality $\lambda\leq\kappa$ as a bound on classical orbital-instability. Whether this relation remains valid away from the near-horizon regime and in more general bh spacetimes has subsequently become an important question.

When a charged particle moves around a charged BH, the Lorentz force acting on the particle support a static equilibrium outside the event horizon. For a fixed spacetime background, the location of the equilibrium is determined by the particle charge-to-mass ratio. By analyzing the stability of such equilibria, Zhao et al. \cite{ZLL} found that the chaos bound is respected in Reissner–Nordstr\"om (RN) and Reissner–Nordstr\"om anti-de Sitter (RN-AdS) spacetimes but can be violated in a variety of other charged BH backgrounds. Their results indicate that the near-horizon relation between the LE and the surface gravity is not universally maintained away from the horizon. When the contribution of particle angular momentum is included, Lei et al. found that the bound can also be violated in RN and RN--AdS spacetimes for suitable BH parameters and sufficiently large particle angular momentum \cite{LG1}. Similar behavior has been reported in other spherically symmetric backgrounds \cite{LG2,LG3,GCYW1,LMX,GCYW2,LMX1}, as well as in Kerr--Newman \cite{KG1}, Kerr--Newman--(A)dS \cite{KG2,KG3}, Kerr--Sen--AdS \cite{KG4,KG5}, and other BH and black-brane geometries \cite{SPN1,LG4,DPS1}. Further studies of the chaos bound can be found in~\cite{LTW1,LTW2,LTW3,LTW4,LTW5,LTW6,LTW7,LTW8,LTW9}. Most of these analyses concern nonspinning charged probes. Related violations have also been discussed in quantum systems \cite{AS,AS1,RRP}, although in some rotating systems the apparent violation can be removed by introducing an appropriate effective temperature \cite{JKY}.

In this paper, we investigate whether the radial LEs of physically admissible spinning charged test particle orbits satisfy the chaos bound in the Kerr--Newman--AdS spacetime. Particular attention is paid to the effects of BH rotation, electromagnetic interaction, AdS curvature, and particle spin. In contrast to spherically symmetric backgrounds, the rotation introduces a preferred axis and gives rise to a pronounced dependence on the relative orientation between the particle total angular momentum and the BH rotation. The particle spin further modifies the radial dynamics through spin--curvature coupling, thereby shifting the boundary between the bound-satisfying and bound-violating regions. The negative cosmological constant modifies both the asymptotic geometry and the effective potential governing the radial motion and can therefore affect the unstable circular orbits as well as the surface gravity. Its influence on the chaos bound is, however, dependent on the background geometry and parameter branch rather than being universally monotonic. Previous studies have examined the chaos bound for spinning particles in spherically symmetric spacetimes \cite{Yang2026,LCL1}. In the present analysis, we additionally restrict the particle parameters to a physically admissible regime. In particular, the spin is constrained so that the particle four-velocity remains timelike, while the charge and other particle parameters are chosen to remain consistent with the test particle approximation. These restrictions exclude parameter regions in which the pole--dipole description or the neglect of particle backreaction would cease to be reliable.

The remainder of this paper is organized as follows. In Sec.~\ref{sec2}, we derive the radial and azimuthal equations of motion for a spinning test particle in the Kerr--Newman--AdS spacetime. In Sec.~\ref{sec3}, we obtain the LE from radial perturbations about unstable circular orbits and examine the chaos bound by analyzing the effects of the particle and BH parameters, with particular attention to the relative orientations of the particle angular momentum, particle spin, and BH rotation. We then investigate the chaos bound behavior in the Kerr--AdS and RN--AdS spacetimes separately. Sec. \ref{sec4} is our discussion and conclusion.

\section{Motions of spinning particles in Kerr–Newman–AdS spacetime}\label{sec2}

\subsection{MPD equations}\label{sec2.1}

In this subsection, we briefly review the dynamics of a charged spinning test particle in a curved spacetime, governed by the Mathisson--Papapetrou--Dixon (MPD) equations \cite{HH1},

\begin{align}
\frac{D p^{\mu}}{D \tau} &= -\frac{1}{2}R^{\mu}_{\nu\alpha\beta}u^{\nu}S^{\alpha\beta} + \tilde{q}F^{\mu}_{\nu}u^{\nu}, \label{eq2.01}\\
\frac{D S^{\mu\nu}}{D \tau} &=  p^{\mu} u^{\nu} - u^{\mu} p^{\nu},
\label{eq2.02}
\end{align}

\noindent where $\frac{D}{D\tau}$ represents the covariant derivative along the particle worldline, parametrized by the proper time $\tau$. The four-momentum and four-velocity of the particle are denoted by $p^{\mu}$ and $u^{\mu}=\frac{dx^{\mu}}{d\tau}$, respectively, and  $\tilde{q}$ is the particle's electric charge. The Riemann curvature tensor is $R^{\mu}_{\nu\alpha\beta}$, and $S_{\mu\nu}$ is an antisymmetric spin tensor, while \(F_{\mu\nu}=\partial_{\mu}A_{\nu}-\partial_{\nu}A_{\mu}\) is the electromagnetic field tensor of the spacetime.

To close the MPD system and fix the representative center-of-mass worldline, we impose the Tulczyjew--Dixon spin supplementary condition (TDSSC) \cite{WT1,WT2},
\begin{eqnarray}
S^{\mu\nu}p_{\nu}=0 .
\label{eq2.03}
\end{eqnarray}
The spin tensor $S^{\mu\nu}$ and the four-momentum $p_{\mu}$ are respectively characterized by the conserved mass $m$ and spin magnitude $\tilde{S}$, defined by

\begin{eqnarray}
m^2 &=&-p^\mu p_\mu, \label{eq2.9}\\
\tilde{S}^2 &=& \frac{1}{2}S_{\mu\nu}S^{\mu\nu}. \label{eq2.8}
\end{eqnarray}

\noindent Introducing the normalized four-momentum, $v^{\mu} = \frac{p^{\mu}}{m}$ and $v^{\mu}v_{\mu} =-1$, its relation to the four-velocity reads \cite{HZ1,HZ2}

\begin{eqnarray}
u^{\mu} - v^{\mu} = \frac{2S^{\mu\nu}v^{\alpha}\left(R_{\nu\alpha\beta\gamma} S^{\beta\gamma}+2\tilde{q}F_{\nu\alpha}\right)}{ S^{\nu\alpha}\left(R_{\nu\alpha\beta\gamma} S^{\beta\gamma}+2\tilde{q}F_{\nu\alpha}\right)+4m^2} .
\label{eq2.04}
\end{eqnarray}

\noindent Thus, owing to the spin-dependent coupling to the background curvature and electromagnetic field, the four-velocity is in general not parallel to the four-momentum.

\subsection{Motion of spinning particles}\label{sec2.3}

The Kerr--Newman--AdS spacetime is a stationary and axisymmetric solution of the Einstein--Maxwell equations describing a rotating charged BH in an asymptotically AdS background with a negative cosmological constant $\Lambda$. Its line element is given by

\begin{eqnarray}
ds^2 &=& -\frac{1}{\rho^2}\left(\Delta_r-a^2\Delta_\theta\sin^2\theta\right)dt^2+\frac{2a\sin^2\theta}{\Xi \rho^2}\left(\Delta_r-(r^2+a^2)\Delta_\theta\right)dtd\phi \nonumber \\
&&+\frac{\sin^2\theta}{\Xi^2\rho^2}\left((r^2+a^2)^2\Delta_\theta-a^2\Delta_r\sin^2\theta\right)d\phi^2+\frac{\rho^2}{\Delta_r	}dr^2+\frac{\rho^2}{\Delta_\theta}d\theta^2,
\label{eq2.1}
\end{eqnarray}

\noindent with the electromagnetic potential 

\begin{eqnarray}
A=-\frac{Qr}{\rho^{2} }(dt-\frac{a\sin^{2}}{\Xi}d\phi).
\label{eq2.2}
\end{eqnarray}

\noindent where

\begin{align}
\rho^{2}    &= r^{2} + a^{2}\cos^{2}\!\theta,\nonumber\\
\Delta_{r}  &= (r^{2}+a^{2})\Bigl(1+\frac{r^{2}}{\ell^{2}}\Bigr)
- 2Mr + Q^{2},\nonumber\\
\Delta_{\theta} &= 1 - \frac{a^{2}}{\ell^{2}}\cos^{2}\!\theta,\nonumber\\
\Xi        &= 1 - \frac{a^{2}}{\ell^{2}}.
\end{align}

\noindent Here $M$, $a$, and $Q$ denote the BH mass, rotation parameter, and electric charge, respectively, while $\ell$ is the AdS curvature radius, related to the cosmological constant by $\Lambda = -3/\ell^{2}$. The surface gravity is 

\begin{eqnarray}
\kappa=\frac{r^2_+-r^2_+(\frac{a^2}{3}+r_+^2)\Lambda -(Q^2+a^2)}{2(r_+^2 +a^2)r_+},
\label{eq2.3}
\end{eqnarray}

\noindent where $r_{+}$ is the event horizon, determined by the largest positive real root of $\Delta_{r}(r)=0$. The rotation and charge of the BH determine the background electromagnetic field, which exerts a Lorentz force on a charged test particle. The surface gravity provides the characteristic horizon scale entering the chaos bound, Eq.~\eqref{eq1.2}, against which the LE associated with the local instability of an unstable circular orbit is compared. In the limits $Q\rightarrow0$ and $a\rightarrow0$, the geometry reduces to the Kerr--AdS and RN--AdS spacetimes, respectively, providing useful reference cases for the subsequent analysis.

The dynamics of spinning test particles in rotating spacetimes have been extensively investigated in Refs.~\cite{HH1,ZHA,HA1,HA2,HH2,HH3}. Following these treatments, we derive the equations of motion for a spinning particle in the Kerr--Newman--AdS spacetime. In the present analysis, the particle motion is restricted to the equatorial plane, while its spin vector is taken to be normal to this plane and hence parallel or antiparallel to the BH rotation axis. Accordingly, $p^{\theta}=0 $ and $S^{\theta\mu}=0$, respectively. These conditions define the symmetry-reduced sector used to construct the radial effective potential. The LE derived below therefore characterizes the growth of radial perturbations while the equatorial configuration and the prescribed spin orientation are maintained. It does not, however, establish the stability of the spin alignment against generic perturbations. A small tilt of the particle spin or a displacement away from the equatorial plane would generally introduce polar motion and spin precession, requiring a coupled perturbation analysis of the full nonequatorial MPD dynamics. Such perturbations are beyond the scope of the present work. Under this restriction, the only independent nonvanishing component of the spin tensor is \(S^{t\phi}\). Substituting this into the TDSSC then gives \cite{HH1}

\begin{eqnarray}
S^{tr}=-\frac{p_\phi}{p_r}S^{t\phi},\qquad	S^{r\phi}=-\frac{p_t}{p_r}S^{t\phi}.
\label{eq2.7}
\end{eqnarray}

\noindent Using Eqs.~\eqref{eq2.7} and \eqref{eq2.8}, we obtain

\begin{eqnarray}
S^{t\phi}
=2\sigma p_r,
\label{eq2.10}
\end{eqnarray}

\noindent where 

\begin{eqnarray}
\sigma=\frac{S}{2\sqrt{g_{rr}\left(g_{t\phi}^{2}-g_{tt}g_{\phi\phi}\right)}},
\label{eq2.11}
\end{eqnarray}

\noindent and $S= \pm\frac{\tilde{S}}{m}$ denotes the signed specific spin of the test particle. The positive (negative) sign corresponds to the spin being aligned (anti-aligned) with the positive $z$ direction. The remaining nonvanishing components of the spin tensor then follow from the Tulczyjew--Dixon supplementary condition, yielding

\begin{eqnarray}
	S^{r\phi} =-2\sigma p_t,\quad	S^{tr}=	-2\sigma p_{\phi}.
	\label{eq2.12}
\end{eqnarray}

The spacetime admits two Killing vector fields, $\xi_{(t)}^a=\left(\frac{\partial}{\partial t}\right)^a$ and $\xi_{(\phi)}^a=\left(\frac{\partial}{\partial \phi}\right)^a$, which generate the stationarity and axisymmetry of the geometry, respectively. The stationary Killing vector $\xi_{(t)}^a$ yields a conserved quantity identified as the energy $\tilde{E}$, given by \cite{HH1}

\begin{eqnarray}
\tilde{E}=mE=-p_t+	\sigma p_t g_{t\phi}^{\prime}-\sigma p_\phi g_{tt}^{\prime} -mqA_t, \label{eq2.13}
\end{eqnarray}

\noindent where $E$ and $q=\frac{\tilde{q}}{m}$ denote the specific energy and specific charge of the particle, respectively, and the prime denotes differentiation with respect to the radial coordinate $r$. The axial Killing vector $\xi_{(\phi)}^a$ is associated with the conserved angular momentum along the $z$-axis, given by \cite{HH1}

\begin{eqnarray}
\tilde{J}=mJ=p_\phi +\sigma p_\phi  g_{t\phi}^{\prime} -\sigma p_t  g_{\phi\phi}^{\prime} +mqA_\phi,
\label{eq2.14}
\end{eqnarray}

\noindent where $J$ is the specific total angular momentum. This conserved quantity contains the orbital, spin-dependent, and electromagnetic contributions to the particle angular momentum. The sign of \(J\) specifies its orientation relative to the symmetry axis: \(J>0\) corresponds to the positive \(z\) direction, whereas \(J<0\) corresponds to the negative \(z\) direction. Solving Eqs.~(\ref{eq2.13}) and (\ref{eq2.14}), we obtain

\begin{eqnarray}
\frac{p_t}{m} &=&-\frac{(1+\sigma g'_{t\phi})\left(E+qA_t\right)	+\sigma g'_{tt}\left(J-qA_\phi\right)}{1-\sigma^2\left((g'_{t\phi})^2-g'_{tt}g'_{\phi\phi}\right)},\label{eq2.17}\\
\frac{p_\phi}{m} &=&\frac{	(1-\sigma g'_{t\phi})\left(J-qA_\phi\right)-\sigma g'_{\phi\phi}\left(E+qA_t\right)}{1-\sigma^2\left((g'_{t\phi})^2-g'_{tt}g'_{\phi\phi}\right)}.	\label{eq2.18}
\end{eqnarray}

\noindent For equatorial motion, $p_{\theta} =0$, and the radial momentum follows from the mass-shell condition. From Eq.~\eqref{eq2.9}, we obtain

\begin{eqnarray}
\frac{p_r^2}{m^2}=g_{rr}\left(-1+\frac{g_{\phi\phi}p_t^2-2g_{t\phi}p_t p_\phi	+g_{tt}p_\phi^2	}{m^2(g_{t\phi}^{2}-g_{tt}g_{\phi\phi})}	\right).
\label{eq2.19}
\end{eqnarray}

From the spin evolution equation, Eq.~\eqref{eq2.02}, we get \cite{HA1,HA2}

\begin{eqnarray}
\frac{DS^{tr}}{D t}&=&p^t\dot r-p^r\nonumber\\
&=&\sigma\left(	R_{\phi t\alpha\beta}S^{\alpha\beta}+	R_{\phi r\alpha\beta}S^{\alpha\beta}\dot r+R_{\phi\phi\alpha\beta}S^{\alpha\beta}\dot\phi\right)+2\tilde{q}\sigma A'_\phi\dot r, \label{eq2.21}\\
\frac{DS^{t\phi}}{D t} 	&=&	p^t\dot\phi-p^\phi \nonumber\\ &=&-\sigma\left(R_{rt\alpha\beta}S^{\alpha\beta}+R_{rr\alpha\beta}S^{\alpha\beta}\dot r	+	R_{r\phi\alpha\beta}S^{\alpha\beta}\dot\phi\right)+ 2\tilde{q}\sigma\left(A'_t+A'_\phi\dot\phi\right).
\label{eq2.22}
\end{eqnarray}

\noindent In the above derivation, the final expressions are obtained by combining the momentum equation, Eq.~\eqref{eq2.01}, with the nonvanishing spin-tensor components in Eq.~\eqref{eq2.12}. We have also used the antisymmetry of the Riemann tensor in its first pair of indices, which implies
$R_{\phi\phi\alpha\beta}=R_{rr\alpha\beta}=0$. Solving Eqs.~\eqref{eq2.21} and \eqref{eq2.22} for the coordinate velocities yields

\begin{eqnarray}
\dot r= \frac{b_1}{a_1}, \label{eq2.23}\\
\dot\phi=\frac{b_2}{a_2}, \label{eq2.24}
\end{eqnarray}

\noindent where

\begin{align}
a_1=&a_2=p^t+\frac{1}{2r^6\Xi}\left(p^t\Xi S^2\left(2\left(4a^2+r^2\right)\left(\Delta_r-a^2\right)-r\left(5a^2+r
\right)\Delta_r'+a^2r^2\Delta_r''\right)\right.
\nonumber\\
&\left.-ap^\phi S^2\left(a^2+r^2\right)\left(8\left(\Delta_r-a^2\right)
-5r\Delta_r'+r^2\Delta_r''\right)+2amqQS\Xi r^3\right),
\label{a1} \\
b_1=&p^r-\frac{p^rS^2\left(2a^2-2\Delta_r+r\Delta_r'\right)}{2r^4
},\label{b1} \\
b_2=&p^\phi-\frac{1}{2r^6}\left(p^\phi S^2\left(2\left(4a^2+3r^2\right)\left(\Delta_r-a^2\right)-r\left(5a^2+4r^2\right)\Delta_r'+r^2\left(a^2+r^2\right)\Delta_r''\right)\right.
\nonumber\\
&\left. -a\Xi p^tS^2\left(8\left(\Delta_r-a^2\right)-5r\Delta_r'+r^2\Delta_r''\right)
-2mqQS\Xi r^3\right).
\label{b2}
\end{align}

\noindent Although the particle mass appears explicitly in the expressions above, by using Eqs.~\eqref{eq2.17} and \eqref{eq2.18}, the explicit mass dependence cancels out. The resulting equations of motion are therefore independent of the particle mass. Since the effective potential governing the orbital-instability is constructed from the radial dynamics, the subsequent analysis focuses on the radial equation of motion. 

\section{Test of the chaos bound in AdS spacetimes}\label{sec3}

The chaos bound in the Kerr--Newman--AdS spacetime has previously been examined using the dynamics of scalar particles, with violations found in certain regions of parameter space \cite{KG2}. In the present work, we extend this analysis to spinning test particles in the same background. Throughout the numerical calculations, we set $M=1$ and consider two spin orientations, with the particle spin directed along or opposite to the positive $z$-axis.

\subsection{Lyapunov exponent}\label{sec3.1}

For a spinless particle, the LE characterizing the local instability of an unstable orbit is determined by the curvature of the effective potential and is therefore directly related to its second radial derivative \cite{CMBWZ}. For charged particles in rotating spacetimes, Refs.~\cite{KG1,KG2} instead employed an effective radial Lagrangian near a local extremum, yielding an LE determined by the potential curvature normalized by the effective kinetic coefficient. For spinning particles, however, spin--curvature coupling introduces additional dynamical degrees of freedom, and the resulting motion cannot in general be reduced to that of a spinless particle. The corresponding instability exponent has been derived in Refs.~\cite{LTW3,CHL}, although those analyses were restricted to static and spherically symmetric black hole backgrounds. By contrast, the radial instability of spinning particles in rotating charged AdS spacetimes, particularly the Kerr--Newman--AdS geometry, has received comparatively limited attention. We therefore derive below the LE associated with radial perturbations of unstable circular orbits in this background. The radial equation of motion can be recast in the effective potential form

\begin{align}
\frac{1}{2}m\dot{r}^2 +\mathcal{V}_{eff}=0,
\label{eq3.1.1}
\end{align}

\noindent where $\mathcal{V}_{eff}$ denotes the effective potential of the spinning particle. Therefore, from the above equation and Eq. \eqref{eq2.23} we get the potential, 

\begin{eqnarray}
\mathcal{V}_{eff}&=&-\frac{m}{2}\left(\frac{b_1}{a_1}\right)^2\nonumber\\
&=& -\frac{m}{2}\left(\frac{	\Xi p^r r^2\mathcal{C}}{\Xi p^t\left(2r^6+S^2\mathcal{A}\right) -ap^\phi S^2\left(a^2+r^2\right)\mathcal{B}	+2amqQS\Xi r^3} \right)^2, \label{eq3.1.2}
\end{eqnarray}

\noindent where 

\begin{align}
\mathcal{A}={}&2\left(4a^2+r^2\right)\left(\Delta_r-a^2\right)-r\left(5a^2+r^2\right)\Delta_r'
+a^2r^2\Delta_r'',\\
\mathcal{B}={}&8\left(\Delta_r-a^2\right)-5r\Delta_r'+r^2\Delta_r'',\\
\mathcal{C}={}&2r^4+2S^2\left(\Delta_r-a^2\right)-rS^2\Delta_r'.
\end{align}

\noindent We consider the regime in which the particle is near an unstable equilibrium orbit located at $r=r_0$. The radial motion is parametrized as $r(t)= r_0 +\epsilon(t)$, and Eq.~\eqref{eq3.1.1} is expanded perturbatively in the small fluctuation $\epsilon$ around $r_0$  yielding

\begin{eqnarray}
\frac{1}{2}\left(m\dot{\epsilon}^2 + \mathcal{V}^{\prime\prime}_{eff}(r_0) \epsilon^2\right) + \mathcal{V}_{eff}(r_0)+ \mathcal{O}(\epsilon) \simeq 0, 
\label{eq3.1.3}
\end{eqnarray}

\noindent where $\mathcal{O}(\epsilon)$ represents higher-order terms in $\epsilon$. For simplicity, we can set $\mathcal{V}_{eff}(r_0)=0$ \cite{CMBWZ}. Neglecting the higher-order contributions, the solution takes the form $\epsilon = c_0e^{\pm \lambda t}$, with $c_0$ an integration constant and the positive sign selected. The quantity $\lambda$ is identified as the LE, and its relation to the effective potential is given by

\begin{eqnarray}
\lambda^2 =-\frac{\mathcal{V}^{\prime\prime}_{eff}(r_0)}{m} =  \frac{1}{2}\frac{d^2}{dr^2}\left(\frac{b_1}{a_1}\right)^2\bigg|_{r = r_0}.\label{eq3.2.3}
\end{eqnarray}

For fixed BH parameters and fixed values of the particle parameters $J$, $S$, and $q$, the conserved energy $E$ and the radius $r_{0}$ of a circular orbit are determined simultaneously by the conditions

\begin{eqnarray}
\mathcal{V}_{eff}\left(r_{0};E\right)=&0, \\
\left. \frac{\partial\mathcal{V}_{eff}\left(r;E\right)}{\partial r}\right|_{r=r_{0}}=&0.
\end{eqnarray}
 
\noindent Here, the dependence of $\mathcal{V}_{eff}$ on the fixed black hole and particle parameters is left implicit for notational simplicity. These two equations are solved simultaneously for $E$ and  $r_{0}$. Since the present analysis focuses on unstable circular orbits, the resulting solutions are further required to satisfy
 
\begin{equation}
\left.\frac{\partial^{2}\mathcal{V}_{eff}(r;E)}{\partial r^{2}}\right|_{r=r_0}<0,
\end{equation}

\noindent which identifies $r_0$ as a local maximum of the radial effective potential and hence corresponds to a radially unstable circular orbit. In the subsequent analysis, we retain only physically admissible solutions lying outside the event horizon, $r_{0}>r_{+}$, and satisfying the constraints associated with the spinning particle dynamics.

\subsection{Kerr–Newman–AdS spacetime}\label{sec3.2}

\begin{figure*}[htbp]
	\centering
	\subcaptionbox{}{\includegraphics[height=0.32\textwidth,width=0.4\textwidth]{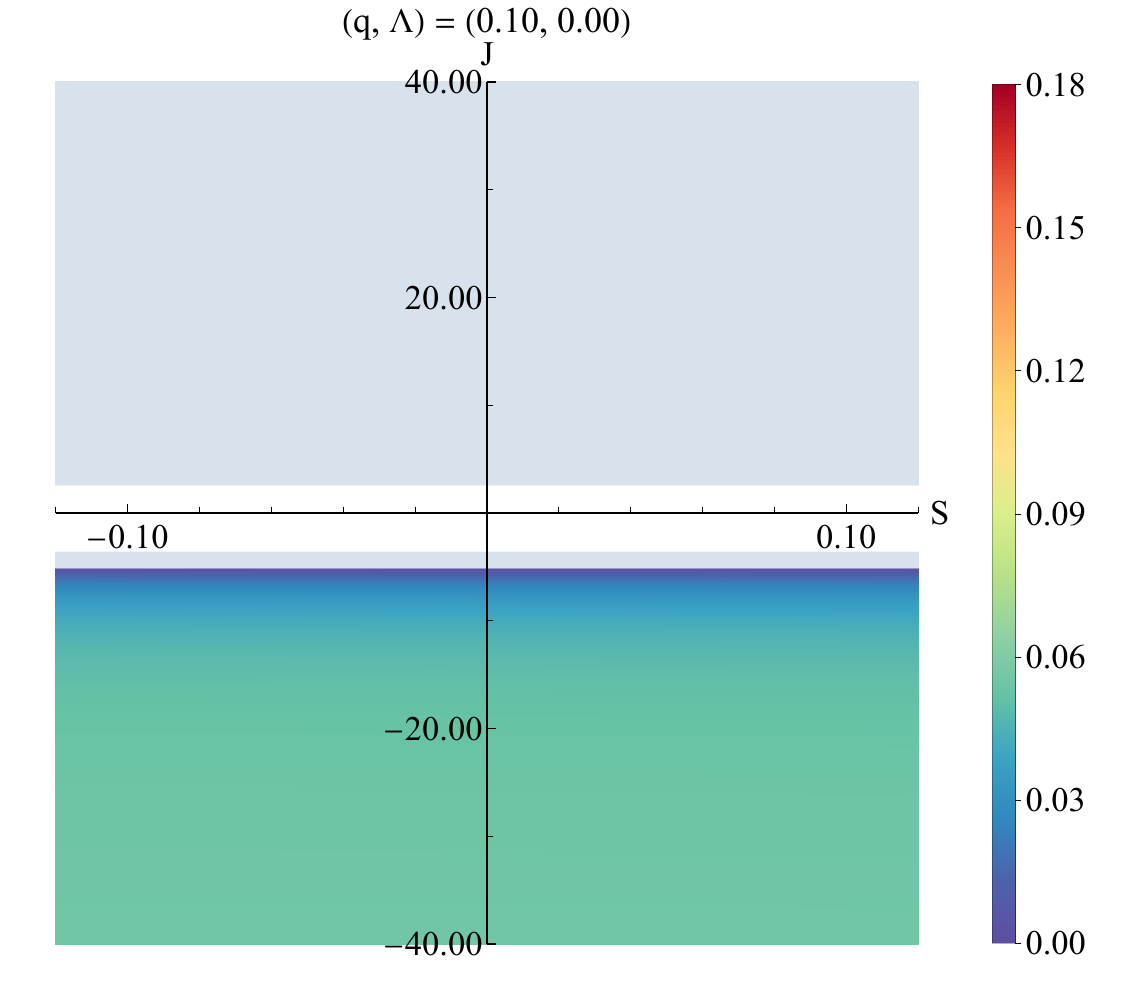}}
	\subcaptionbox{}{\includegraphics[height=0.32\textwidth,width=0.4\textwidth]{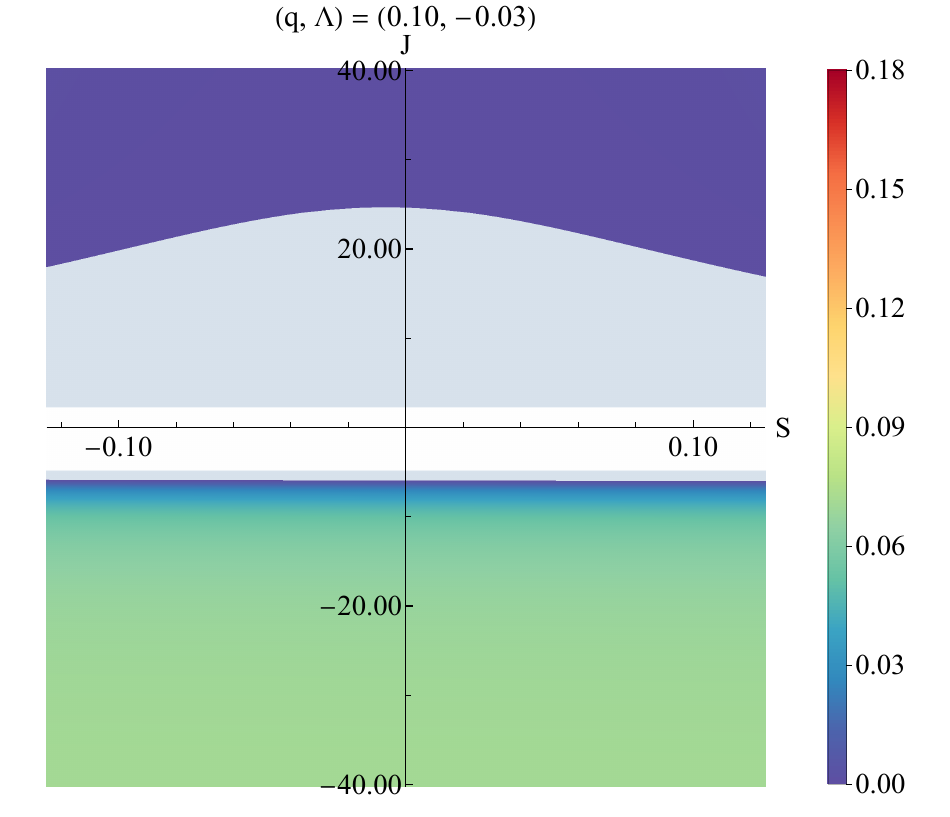}}
	\subcaptionbox{}{\includegraphics[height=0.32\textwidth,width=0.4\textwidth]{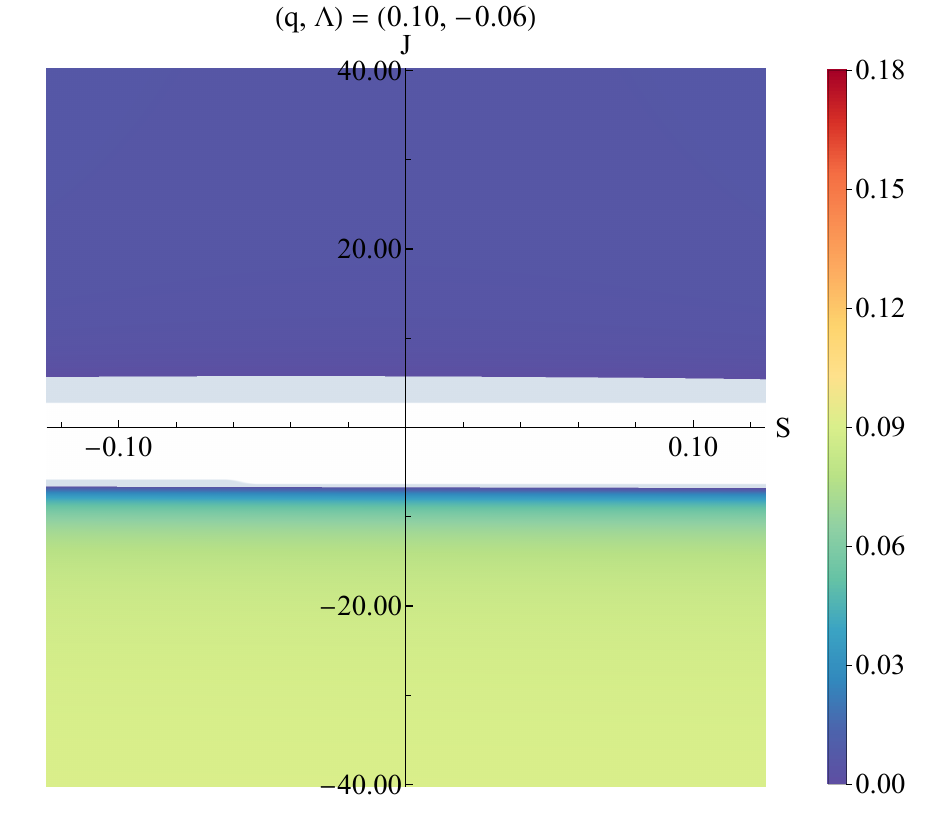}}
	\subcaptionbox{}{\includegraphics[height=0.32\textwidth,width=0.4\textwidth]{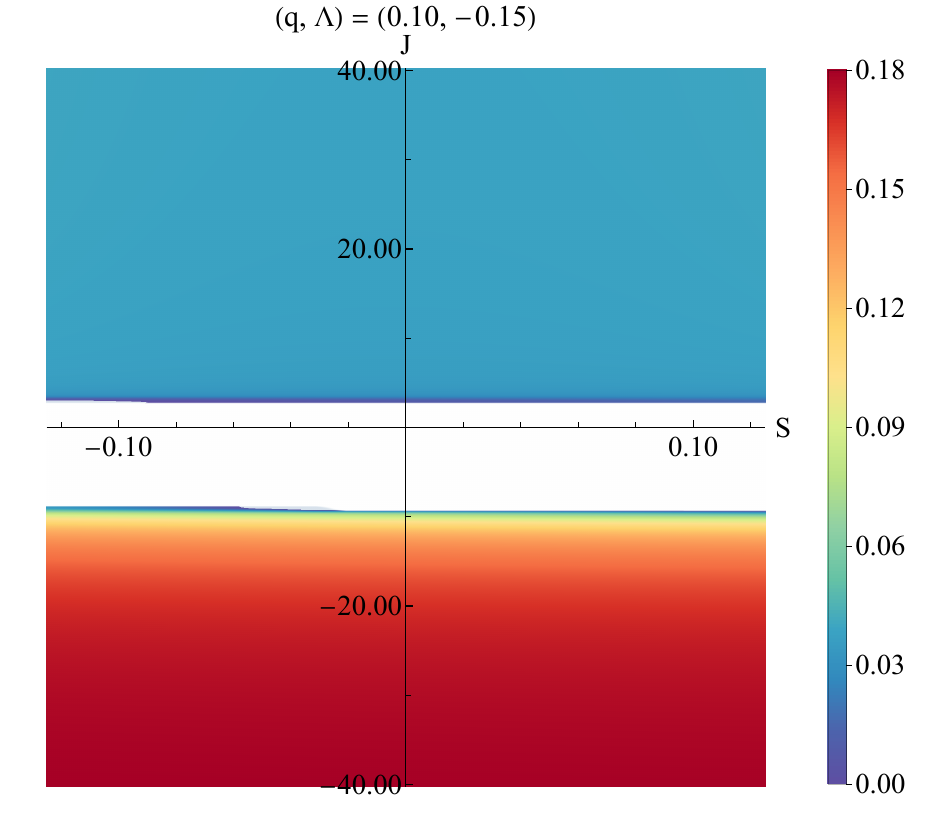}}\\[5mm]
	\subcaptionbox{}{\includegraphics[height=0.32\textwidth,width=0.4\textwidth]{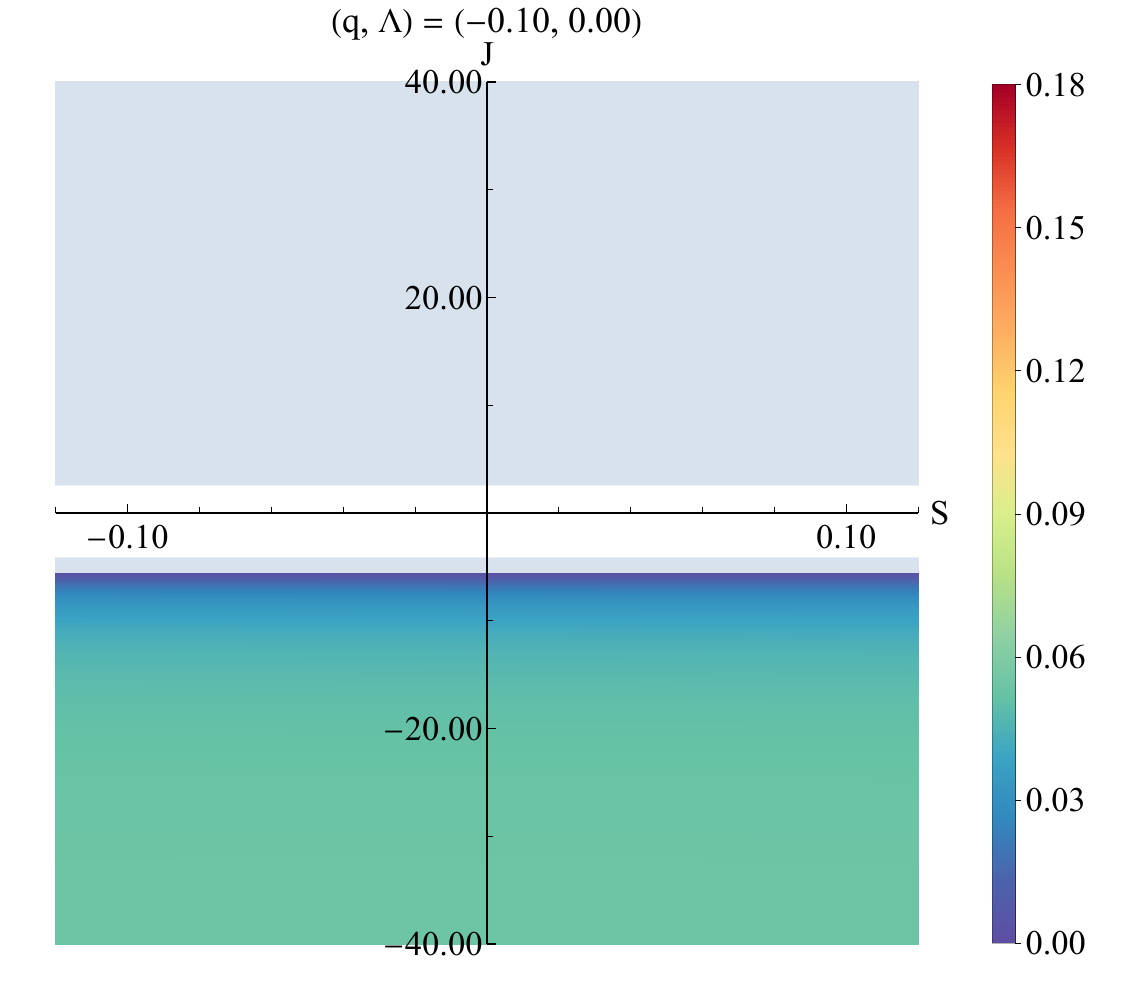}}
	\subcaptionbox{}{\includegraphics[height=0.32\textwidth,width=0.4\textwidth]{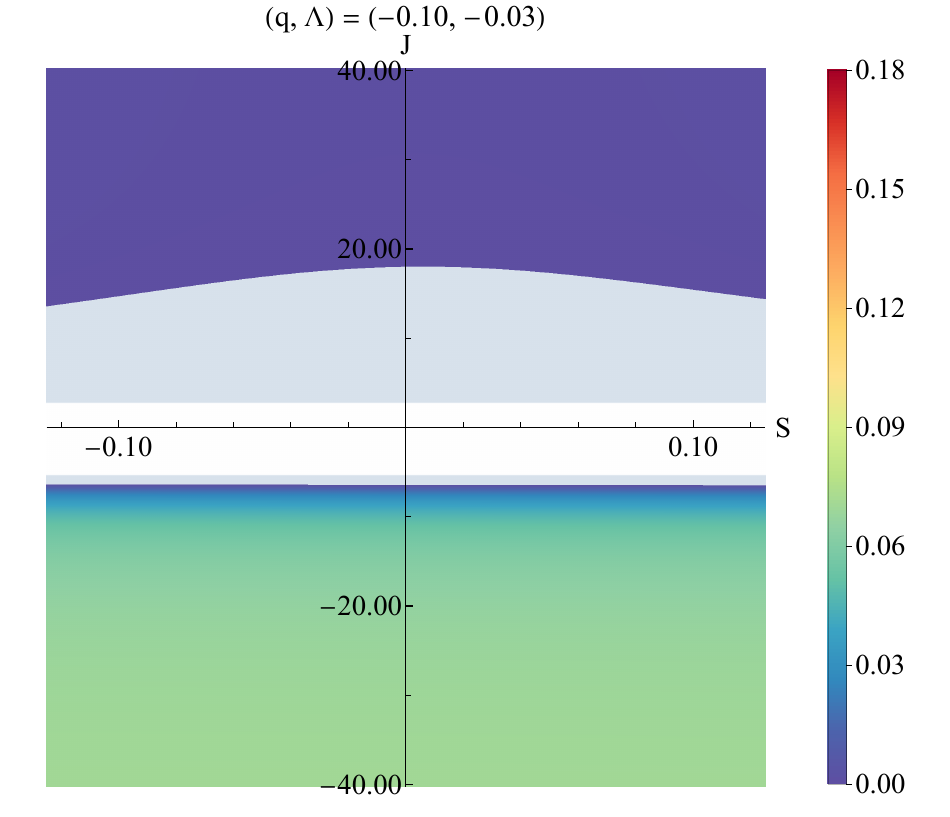}}
	\subcaptionbox{}{\includegraphics[height=0.32\textwidth,width=0.4\textwidth]{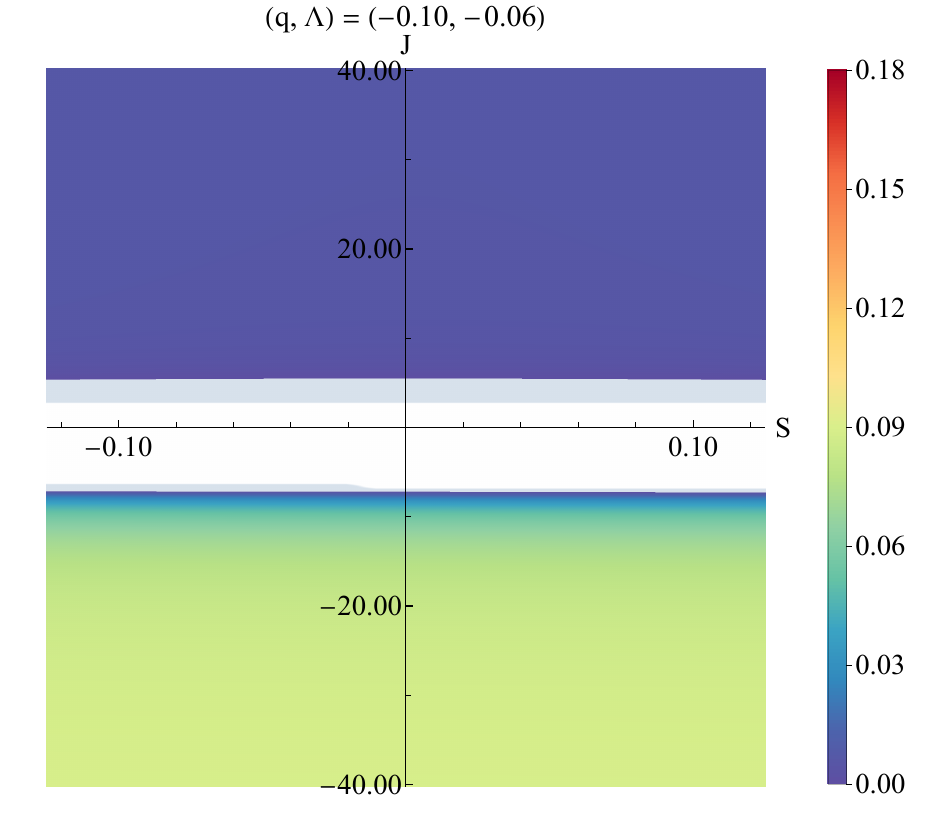}}
	\subcaptionbox{}{\includegraphics[height=0.32\textwidth,width=0.4\textwidth]{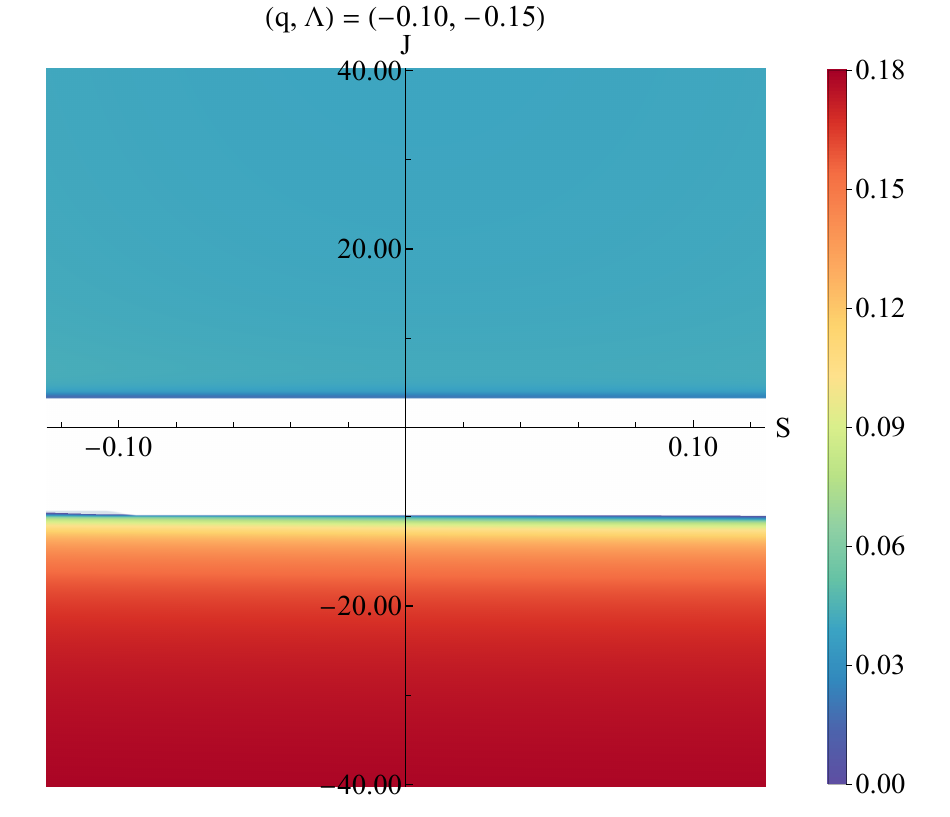}}
	\caption{Dependence of the difference between the LE and surface gravity on the particle spin and total angular momentum for $a=0.365$ and $Q=0.90$.}
	\label{Fig1}
\end{figure*}

\begin{figure*}[htbp]
	\centering
	\subcaptionbox{}{\includegraphics[width=0.32\textwidth]{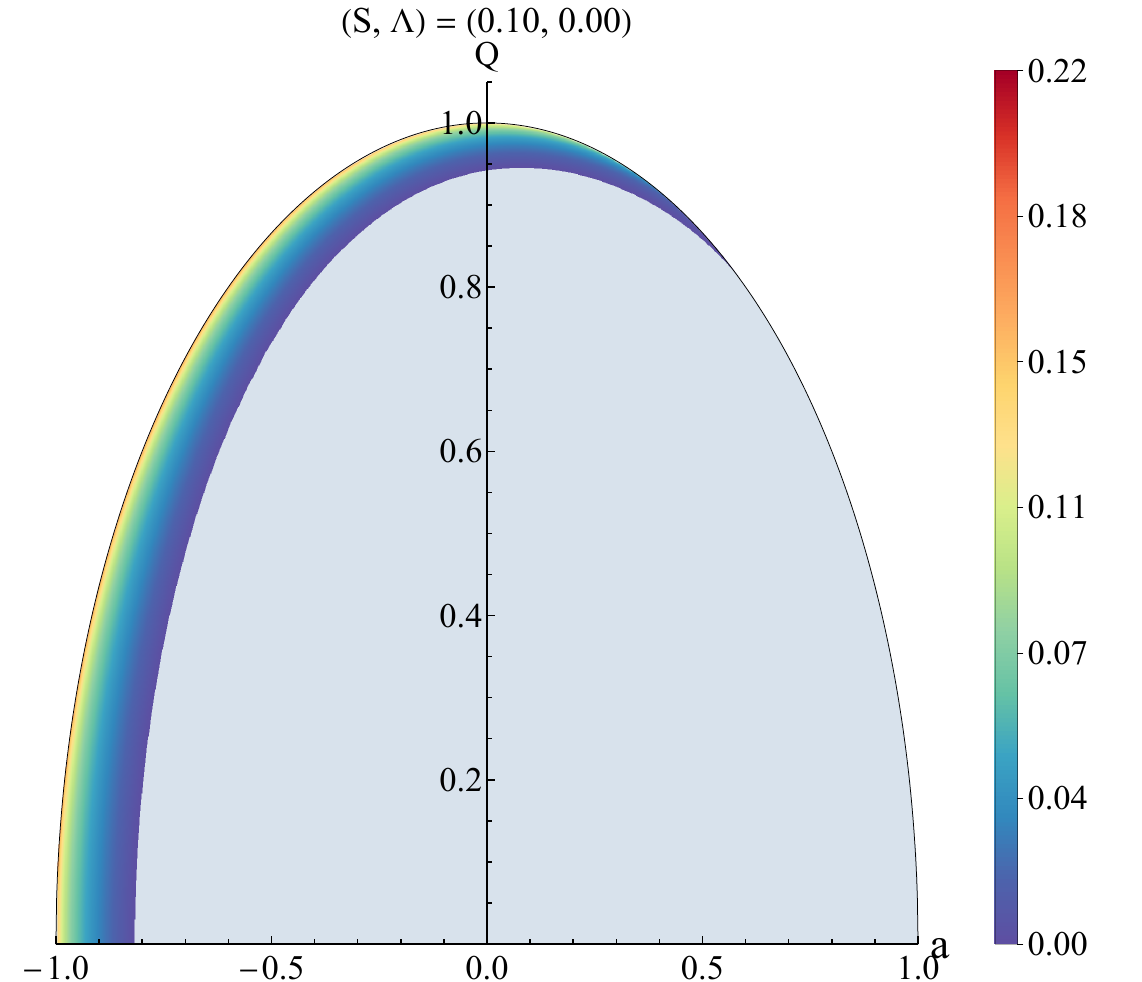}}
	\subcaptionbox{}{\includegraphics[width=0.32\textwidth]{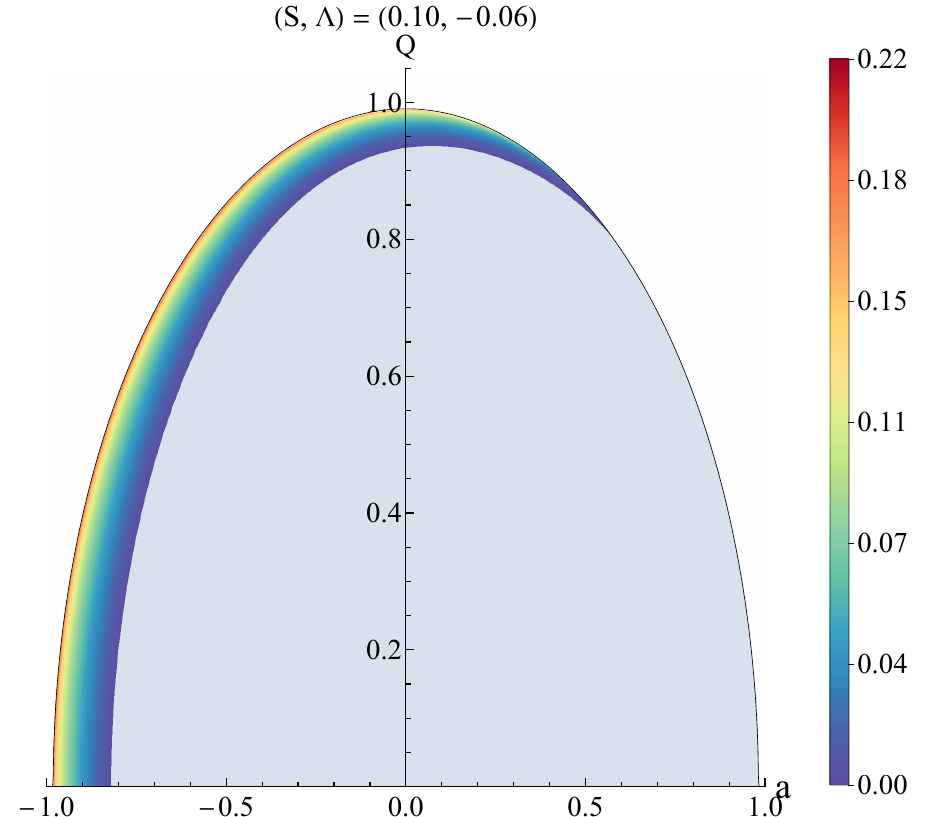}}
	\subcaptionbox{}{\includegraphics[width=0.32\textwidth]{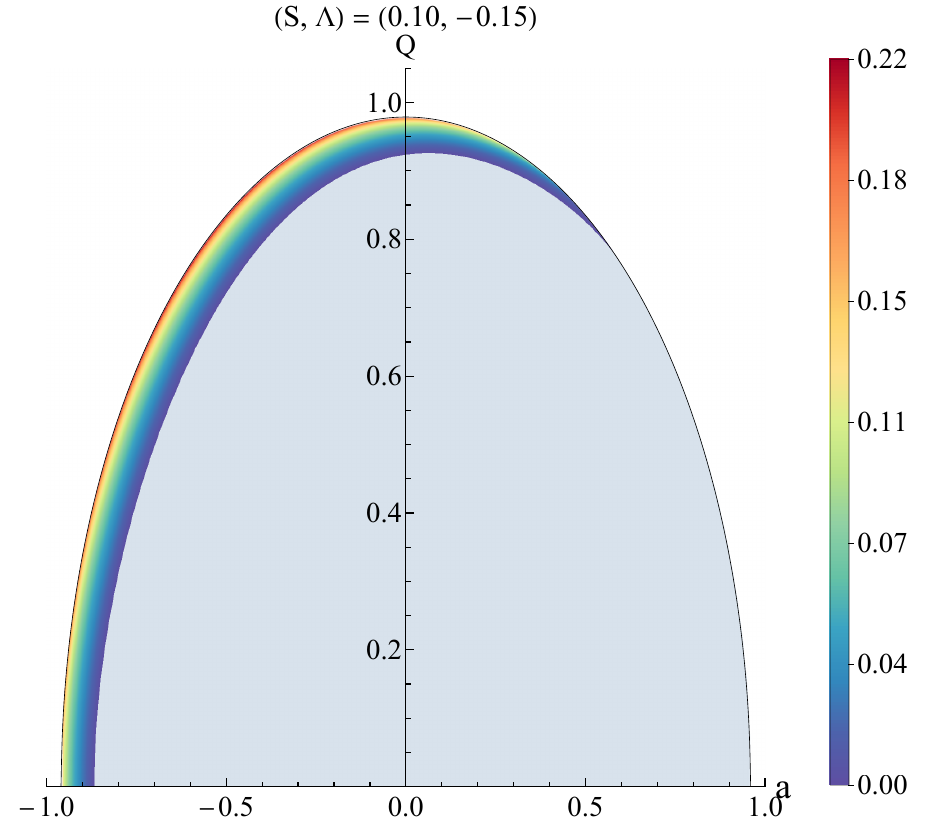}}\\[5mm]
	\subcaptionbox{}{\includegraphics[width=0.32\textwidth]{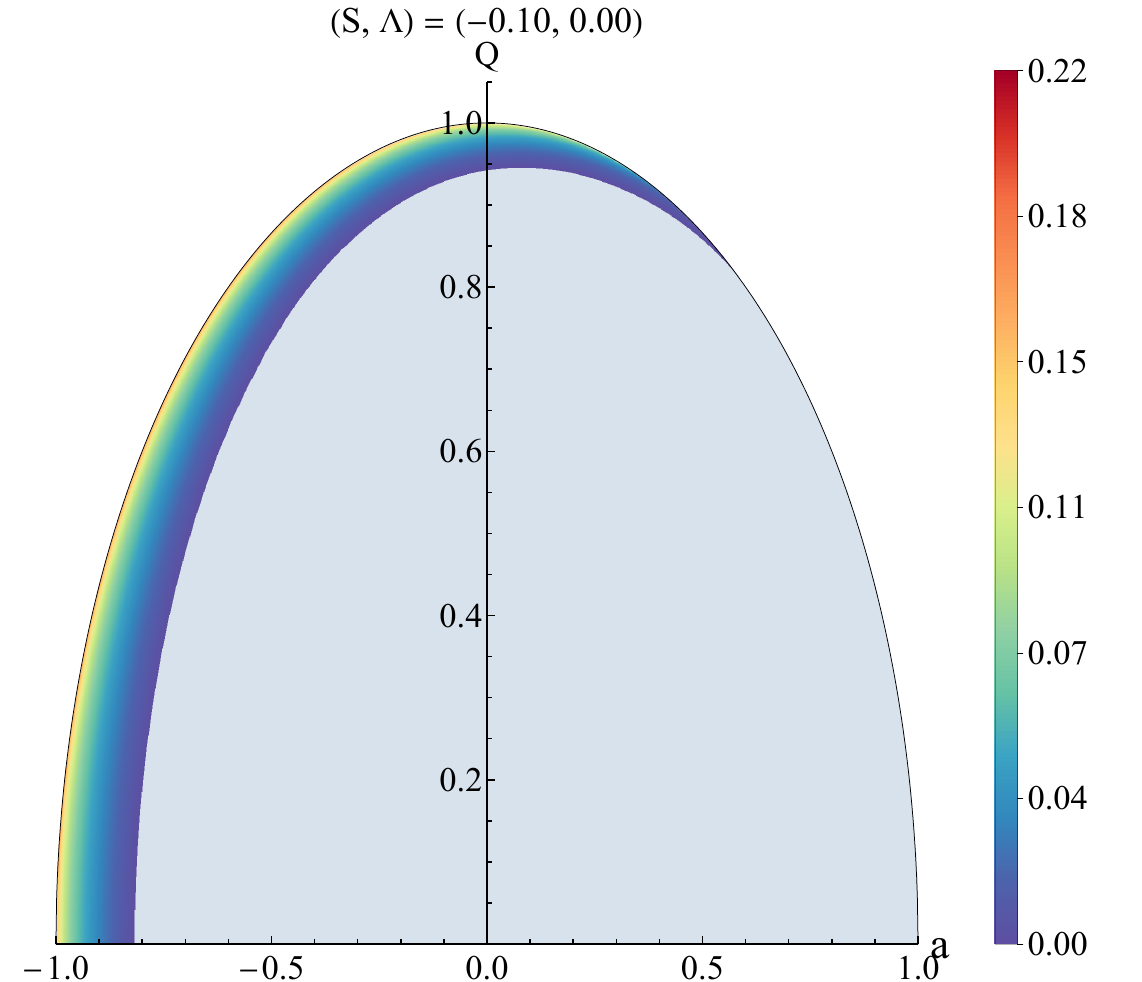}}
	\subcaptionbox{}{\includegraphics[width=0.32\textwidth]{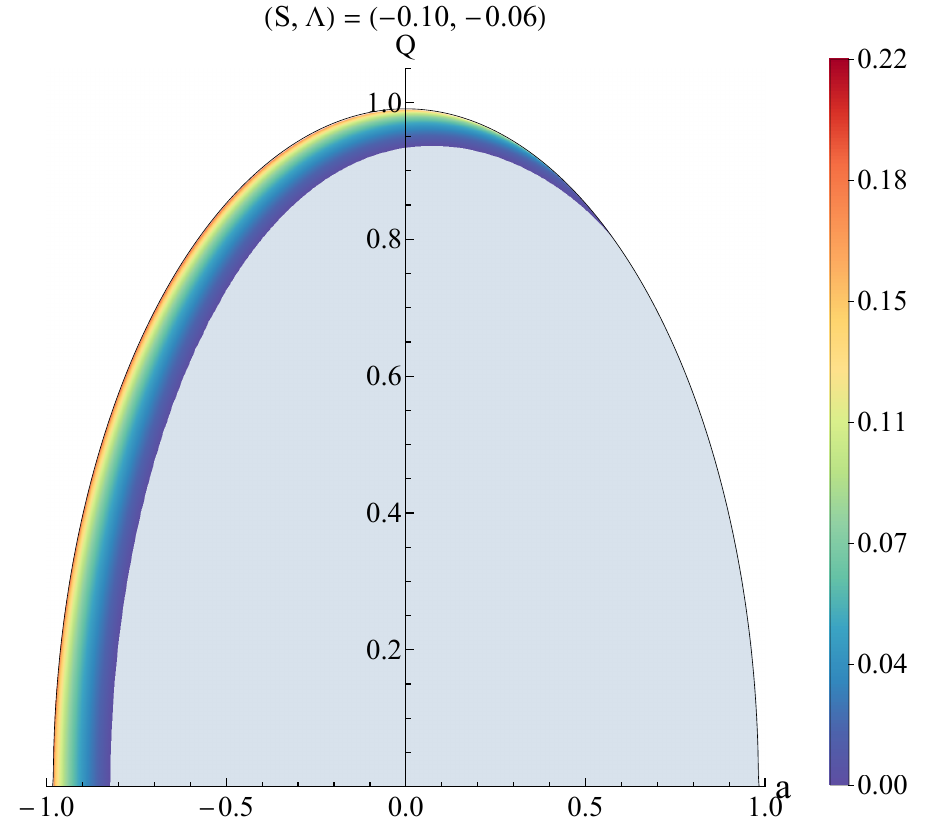}}
	\subcaptionbox{}{\includegraphics[width=0.32\textwidth]{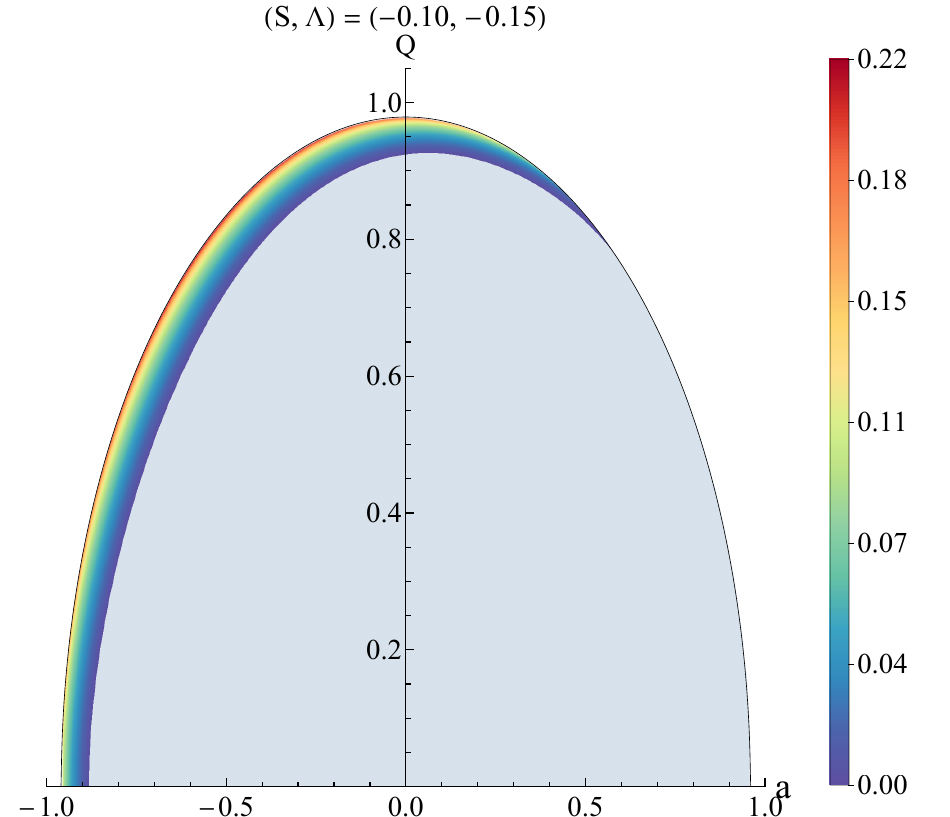}}
	\caption{Dependence of the difference between the LE and surface gravity on the BH charge and rotation parameter for $q=0.10$ and $J=25.00$.}
	\label{Fig2}
\end{figure*}

\begin{figure*}[htbp]
	\centering
	\subcaptionbox{}{\includegraphics[width=0.32\textwidth]{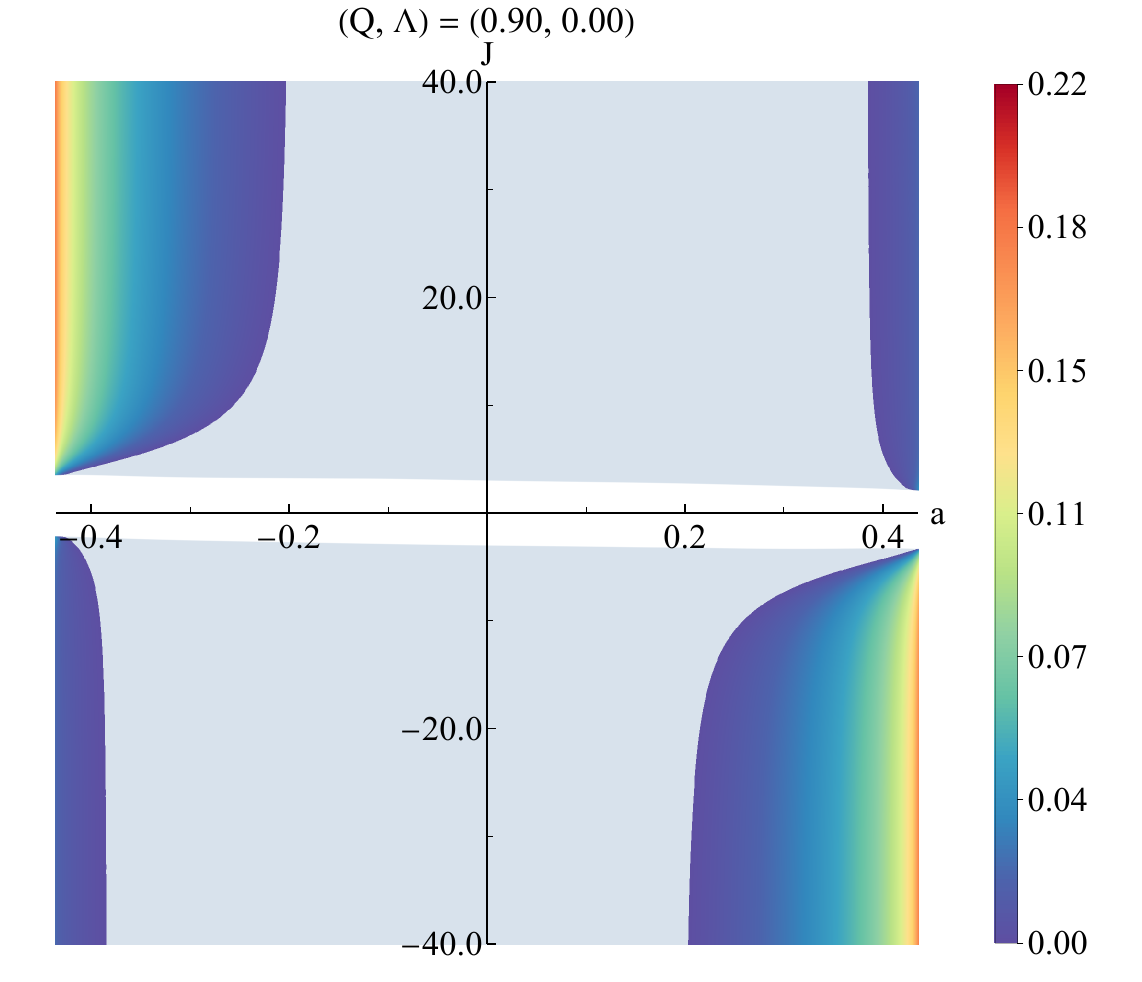}}
	\subcaptionbox{}{\includegraphics[width=0.32\textwidth]{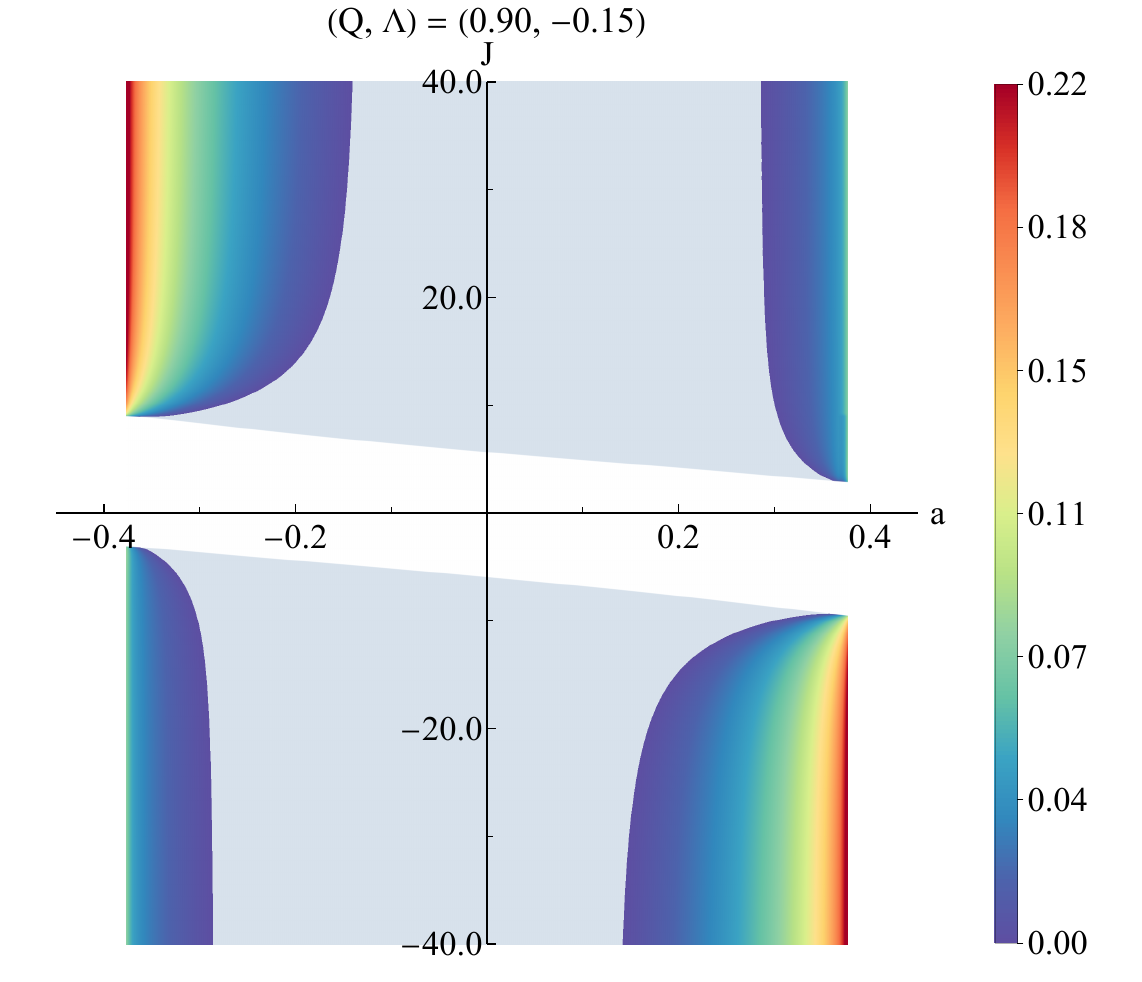}}
	\subcaptionbox{}{\includegraphics[width=0.32\textwidth]{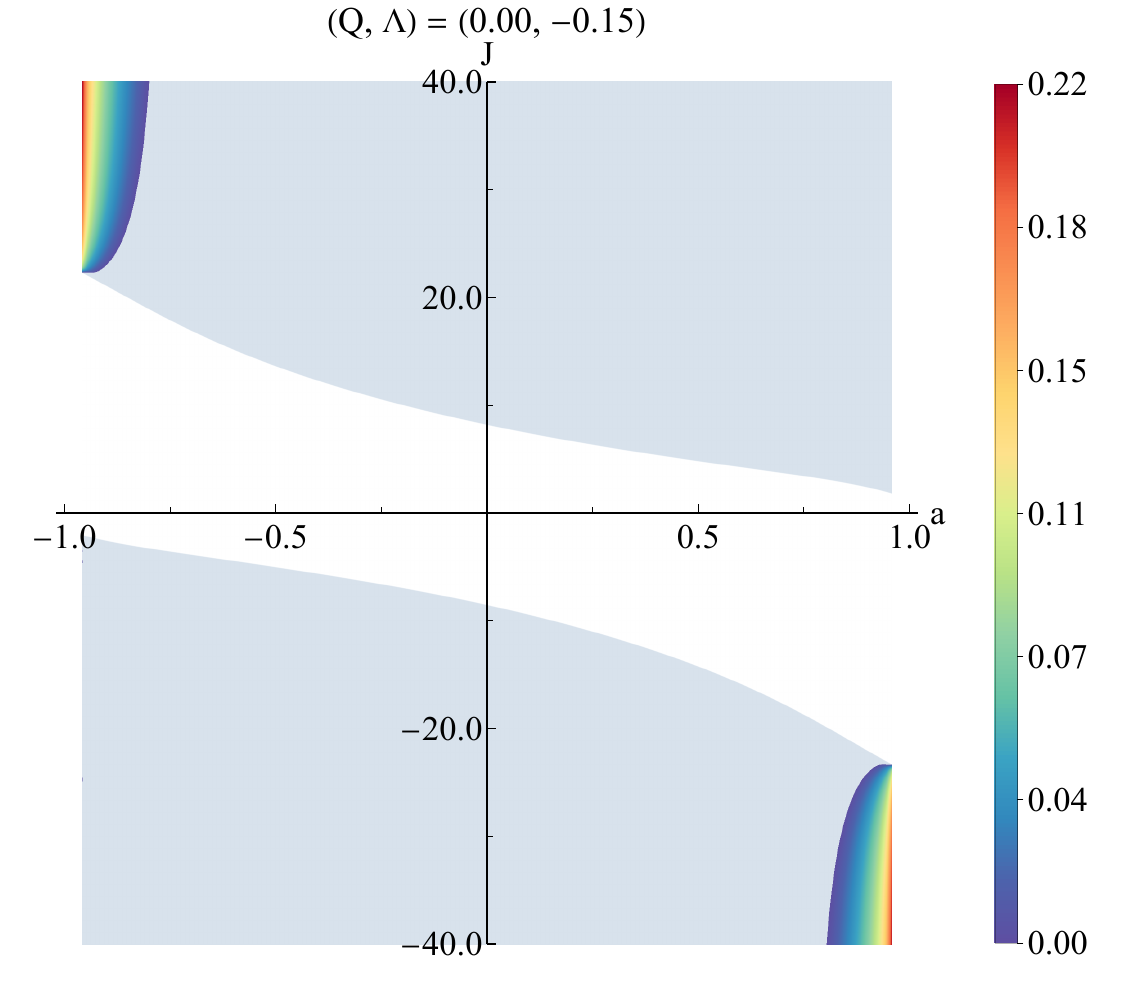}}
	\caption{Dependence of the difference between the LE and surface gravity on the BH rotation parameter and the particle total angular momentum for $S=0.10$ and $q=0.10$.}
	\label{Fig3}
\end{figure*}

\begin{figure}[htbp]
	\centering
	\setlength{\tabcolsep}{2pt}
	\includegraphics[width=0.5\textwidth]{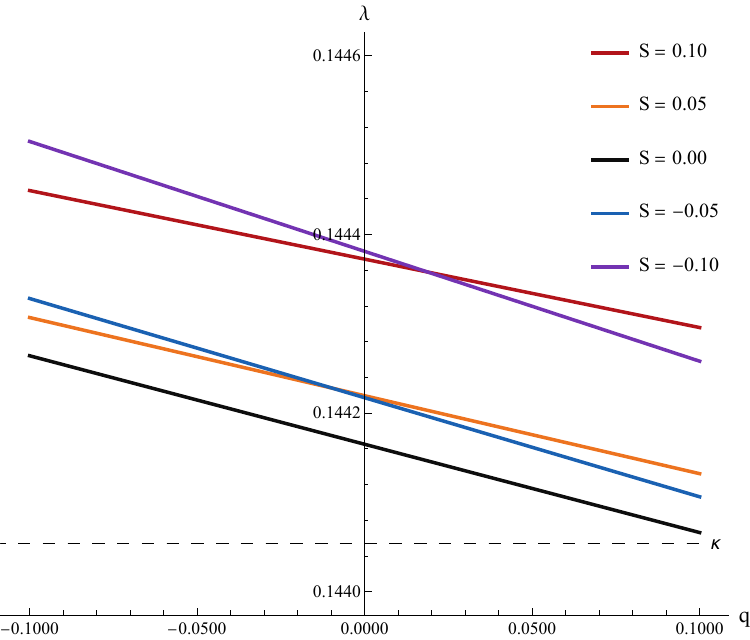}
	\caption{The LEs are shown as a function of particle charge for $Q=0.90$, $a=0.365$, $\Lambda =-0.03$ and $J=25.00$. }
	\label{Fig4}
\end{figure}

\begin{figure}[htbp]
	\centering
	\setlength{\tabcolsep}{2pt}
	\includegraphics[width=0.5\textwidth]{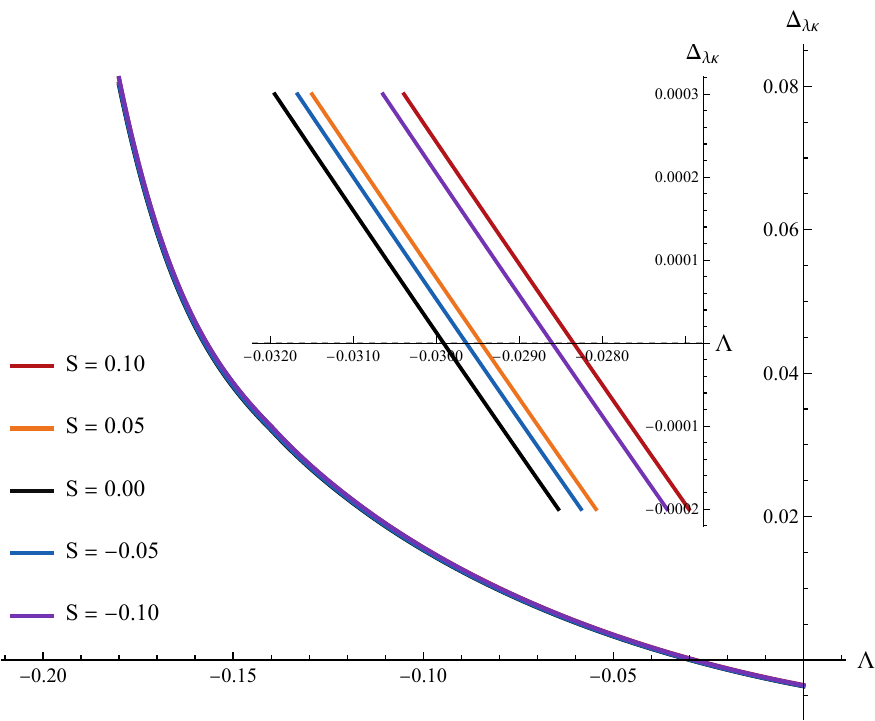}
	\caption{The LEs are shown as a function of the cosmological constant for $Q=0.90$, $a=0.365$, $J=25.00$ and $q =0.10$.}
	\label{Fig5}
\end{figure}

Here, using Eqs.~\eqref{eq3.2.3} and \eqref{eq2.3}, we numerically investigate the relation between the radial LE and the surface gravity in the Kerr--Newman--AdS spacetime. The results are presented in Figs.~\ref{Fig1}--\ref{Fig5}, where the condition $\lambda\leq\kappa$ is examined over the parameter ranges considered.. Without loss of generality, we set the BH mass to $M=1$ throughout the numerical analysis.

We first examine the effects of the particle spin and total angular momentum on chaos bound violation in Fig.~\ref{Fig1}. The figure consists of eight panels, with the particle spin shown along the horizontal axis and the total angular momentum along the vertical axis. The gray regions denote parameter values for which the chaos bound is satisfied, $\lambda\leq\kappa$, whereas the white regions correspond to cases in which no unstable circular orbit exists. The color scale represents the deviation $\Delta_{\lambda\kappa}=\lambda-\kappa$. The same conventions are adopted in the subsequent figures.

Panel~(a) corresponds to a vanishing cosmological constant, $\Lambda=0$, for which the background reduces to the Kerr--Newman geometry. On the positive-$J$ branch, where the particle total angular momentum is aligned with the fixed positive BH rotation, the chaos bound remains satisfied throughout the parameter range considered. On the negative-$J$ branch, where the  total angular momentum is oppositely directed relative to the BH rotation, violation occurs once $|J|$ exceeds a critical value. The location of this boundary depends only weakly on the particle spin, with variations in the spin producing only modest changes in $\Delta_{\lambda\kappa}$ at fixed angular momentum. Panel~(e), obtained by reversing the particle charge to $q=-0.10$, displays the similar qualitative structure. When the cosmological constant is decreased to $\Lambda=-0.03$, as shown in panel~(b), a violating region also emerges on the positive-$J$ branch, although it remains smaller than that on the negative-$J$ branch. The magnitude of the violation, quantified by $\Delta_{\lambda\kappa}$, is also generally smaller on the positive-$J$ side. On this branch, the critical magnitude $|J_c|$ reaches its maximum near $S=0$ and decreases as $|S|$ increases, whereas its dependence on the particle spin is considerably weaker on the negative-$J$ branch.	As the cosmological constant is further decreased to $\Lambda=-0.06$ and $-0.15$, as shown in panels~(c) and~(d), the violation boundary shifts in a clearly branch-dependent manner. On the positive-$J$ branch, the critical magnitude $|J_c|$ decreases, so that violation can occur at smaller values of the particle total angular momentum. By contrast, on the negative-$J$ branch, $|J_c|$ increases, shifting the onset of violation toward larger angular momentum magnitudes. Thus, a more negative cosmological constant does not uniformly lower the angular momentum threshold. Instead, it affects the two angular momentum branches differently.
 
In panels~(e)--(h), the particle charge is reversed from $q=0.10$ to $q=-0.10$.  For $\Lambda=-0.03$, comparison of panels~(b) and~(f) shows that reversing the sign of the charge affects the two angular momentum branches differently. On the positive-$J$ branch, the bound-satisfying region is reduced and the critical magnitude required for violation decreases. The threshold remains largest near vanishing spin and decreases as the spin moves away from zero in either direction. On the negative-$J$ branch, by contrast, the charge reversal produces a slight increase in $|J_c|$, although the shift is comparatively small. The same qualitative asymmetry persists for more negative values of the cosmological constant. In particular, panel~(h) shows only a modest displacement of the violation boundary relative to panel~(d). Within the parameter range considered, reversing the sign of the particle charge tends to favor violation on the positive-$J$ branch, while slightly increasing the critical angular momentum for its onset on the negative-$J$ branch.

These results demonstrate that the response of the chaos bound to the cosmological constant is branch dependent within the parameter range considered. The negative-$J$ branch generally exhibits a broader bound-violating region than the positive-$J$ branch; however, this does not imply that making the cosmological constant more negative uniformly lowers the critical angular momentum on both branches. For the fixed positive BH rotation considered here, as the cosmological constant becomes more negative, the critical magnitude decreases on the positive-$J$ branch but increases on the negative-$J$ branch. This asymmetric shift reflects the fact that varying the cosmological constant modifies both the unstable circular orbit, and hence its local instability scale, and the BH surface gravity. The violation boundary is therefore determined by the relative variation of these two scales, rather than by a monotonic enhancement or suppression of either quantity alone. Reversing the particle charge further modifies this balance through the electromagnetic interaction, producing an additional branch-dependent shift of the boundary. The structure of Fig.~\ref{Fig1} thus reflects the interplay among the BH rotation, particle angular momentum and spin, electromagnetic coupling  and the AdS geometry.

Figure~\ref{Fig2} illustrates the combined effects of the BH charge and rotation parameter on chaos bound violation. In panel~(a), the cosmological constant is set to $\Lambda=0$, so that the background spacetime reduces to the Kerr--Newman geometry. The gray region, where the bound is satisfied, occupies most of the parameter plane, whereas violation is confined to a relatively limited region. In the nonrotating limit ($a=0$), the spacetime reduces to the RN geometry, and violation occurs only within a finite interval of the BH charge. As the rotation parameter increases toward positive values, this interval gradually narrows and eventually disappears at sufficiently large positive $a$. By contrast, as the rotation parameter becomes more negative, the charge interval over which the bound is violated becomes broader. For a fixed BH charge and a fixed $|a|$, with the remaining parameters held fixed, if the bound is satisfied when the BH rotation is anti-aligned with the particle angular momentum, it is also satisfied in the aligned configuration, whereas the converse does not necessarily hold within the parameter range considered. This directional asymmetry reflects the dependence of the orbital dynamics on the relative orientation between the particle angular momentum and the BH rotation. The BH charge, on the other hand, modifies both the spacetime geometry and the electromagnetic interaction experienced by the charged particle, and therefore affects both the unstable orbit dynamics and the horizon properties. 

Keeping the particle spin fixed and making the cosmological constant more negative, with $\Lambda=-0.06$ and $-0.15$ in panels~(b) and~(c), respectively, reduces both the maximum charge and the range of rotation parameters over which the bound is violated. Consequently, the violating region progressively contracts as the cosmological constant becomes more negative. For a fixed cosmological constant, reversing the spin orientation, as shown in panels~(d)--(f), does not qualitatively alter the structure of the violating region. The comparatively weak dependence on the spin orientation indicates that, within the small-spin regime considered here, spin--curvature coupling induces only modest shifts in the violation boundary relative to the more pronounced variations associated with the background parameters and particle angular momentum.

These results reveal distinct effects of the BH rotation and charge on the relation between the orbital-instability scale and the surface gravity scale. When the BH rotation is aligned with the particle angular momentum, increasing its magnitude of the rotation tends to suppress chaos bound violation, whereas increasing the rotation magnitude in the anti-aligned direction tends to favor it within the parameter range considered. In the static limit, violation occurs only within a finite interval of the BH charge, indicating that the charge modifies both the location and extent of the violating region through its effects on the background geometry and electromagnetic interaction. Making the cosmological constant more negative further contracts this region. This behavior reflects simultaneous changes in the unstable circular orbit and its LE, as well as in the BH surface gravity, rather than a direct suppression of chaos by the AdS background itself. Within the parameter regime explored here, the violation therefore reflects the interplay among the background geometry, BH rotation, and charge-dependent geometric and electromagnetic effects, while the particle spin induces only comparatively modest shifts in the violation boundary.

Figure~\ref{Fig3} illustrates the combined effects of the particle total angular momentum and the BH rotation parameter on chaos bound violation. Since both the rotation parameter and total angular momentum are varied, the onset of violation is described by a boundary in the $(a,J)$ plane rather than by a single critical value of either parameter. In panel~(a), the BH charge is fixed at $Q=0.90$ and the cosmological constant is set to $\Lambda=0$. The shape of the violation boundary depends strongly on the relative orientation of the particle angular momentum and the BH rotation. At fixed rotation, the critical magnitude $|J_c|$ differs between the aligned and anti-aligned angular momentum branches. Conversely, at fixed total angular momentum, the aligned branch generally requires a larger critical rotation magnitude $|a_c|$ for the violation. The aligned branch also occupies a larger portion of the $(a,J)$ plane in which the chaos bound is satisfied, indicating that the orientation dependence cannot be characterized by a single critical threshold but is instead reflected in the full two-dimensional structure of the violation boundary.

When the cosmological constant becomes more negative, from $\Lambda=0$ to $\Lambda=-0.15$, with the charge held fixed, as shown in panel~(b), the physically allowed range of the rotation parameter narrows, and the violation boundary shifts appreciably. Compared with panel~(a), the bound-satisfying region within the corresponding physically allowed parameter domain is overall reduced, while the aligned branch continues to occupy a larger portion of the parameter plane in which the bound is satisfied. Owing to the nontrivial shape of the boundary, however, the shift in the critical angular momentum or rotation parameter depends on the particular slice through the $(a,J)$ plane. The shift of the boundary reflects the simultaneous influence of the AdS geometry on the unstable circular orbit and the surface gravity, and hence on the relative balance between the local orbital-instability and surface-gravity scales.

Panel~(c) corresponds to $Q=0$, for which the background reduces to Kerr--AdS and the electromagnetic interaction between the particle and the BH is absent. The violation boundary is  substantially modified. In particular, for positive values of the BH rotation parameter, the chaos bound remains satisfied throughout the branch in which the particle angular momentum is aligned with the BH rotation over the parameter range shown, whereas violation persists in part of the angular-momentum anti aligned branch. Compared with panel~(b), however, the bound-violating region on the anti-aligned branch becomes smaller, with the boundary shifted toward larger magnitudes of the particle angular momentum along the corresponding fixed-$a$ slices. Thus, removing the charge reduces the extent of the bound-violating region on this branch within the parameter range considered, but does not eliminate the violation.  Comparison among the three panels shows that the BH rotation, cosmological constant, and charge modify the violation boundary in different ways, with the resulting behavior depending sensitively on the relative orientation of the particle total angular momentum and the BH rotation.

Figure~\ref{Fig4} shows the dependence of the LE on the particle charge. For all spin configurations considered, the exponent decreases monotonically as the particle charge $q$ increases in the positive direction. Throughout the range displayed, the exponent remains larger than the surface gravity, and the chaos bound is therefore violated. This behavior occurs for the particular set of relatively large particle angular momentum, BH charge, and rotation parameter adopted here. In addition, the LEs of the spinning particles are systematically larger than that of the spinless particle. At smaller charge, the curves for $S=-0.05$ and $S=-0.10$ lie above those for $S=0.05$ and $S=0.10$, respectively. Each pair intersects at a finite value of the particle charge, beyond which this ordering is reversed. The monotonic decrease of the LE with increasing positive particle charge reflects the modification of the local dynamics of the unstable circular orbit by the electromagnetic interaction. For particle and BH charges of the same sign, increasing the particle charge strengthens the electromagnetic repulsion and changes the location and local properties of the unstable circular orbit; within the parameter range considered, these changes are accompanied by a reduction in its radial instability. The crossings between curves with opposite spin orientations further show that the influence of spin--curvature coupling is sensitive to the particle charge. As the particle charge varies, the interplay between the electromagnetic interaction and spin--curvature coupling affects the local orbital instability differently for the two spin orientations, leading to the observed reversal in the ordering of the LEs.

Figure~\ref{Fig5} shows the dependence of chaos bound violation on the cosmological constant. For all spin configurations considered, the deviation $\Delta_{\lambda\kappa}$ increases as the cosmological constant becomes more negative. Once $\Delta_{\lambda\kappa}$ crosses zero, the local LE exceeds the surface gravity scale, and the bound is violated. Different spin configurations exhibit distinct critical values of the cosmological constant at which this threshold crossing occurs, indicating that the spin shifts the onset of violation. The location of this threshold varies with both the magnitude and orientation of the particle spin. These results show that, within the parameter range considered, making the cosmological constant more negative shifts the relative balance between the local orbital-instability and surface gravity scales toward the bound-violating regime, while spin--curvature coupling introduces an additional spin-dependent shift in the violation threshold.

\subsection{Kerr–AdS spacetime}\label{sec3.3}

\begin{figure*}[htbp]
	\centering
	\subcaptionbox{}{\includegraphics[width=0.4\textwidth]{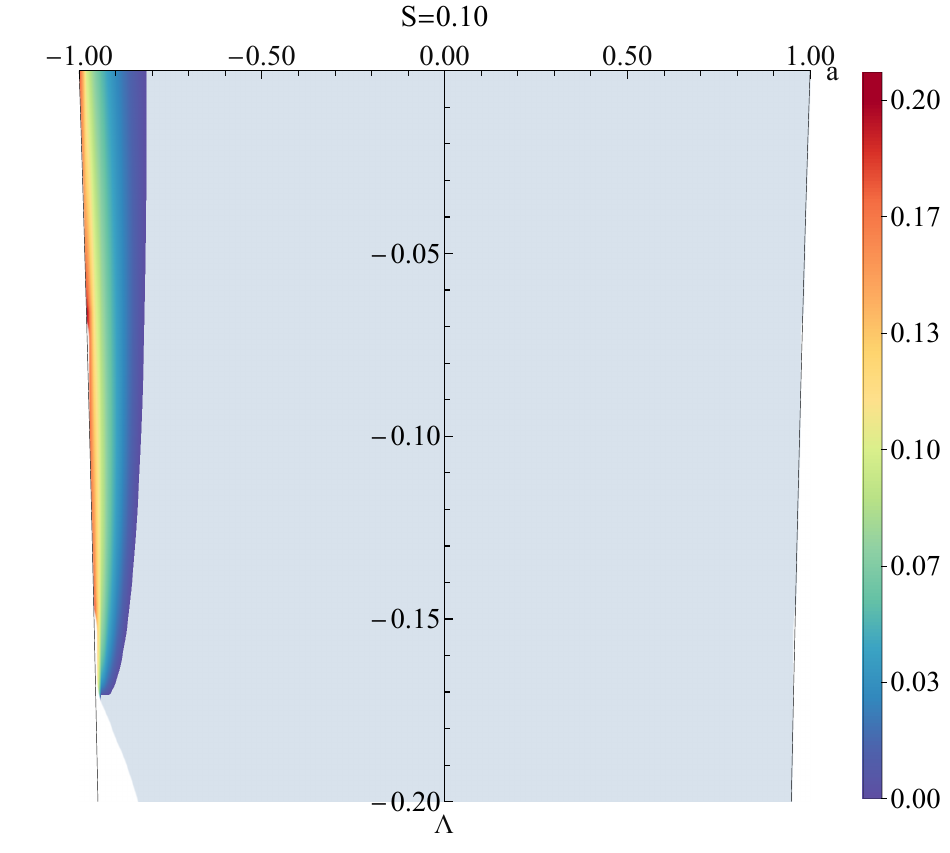}}
	\subcaptionbox{}{\includegraphics[width=0.4\textwidth]{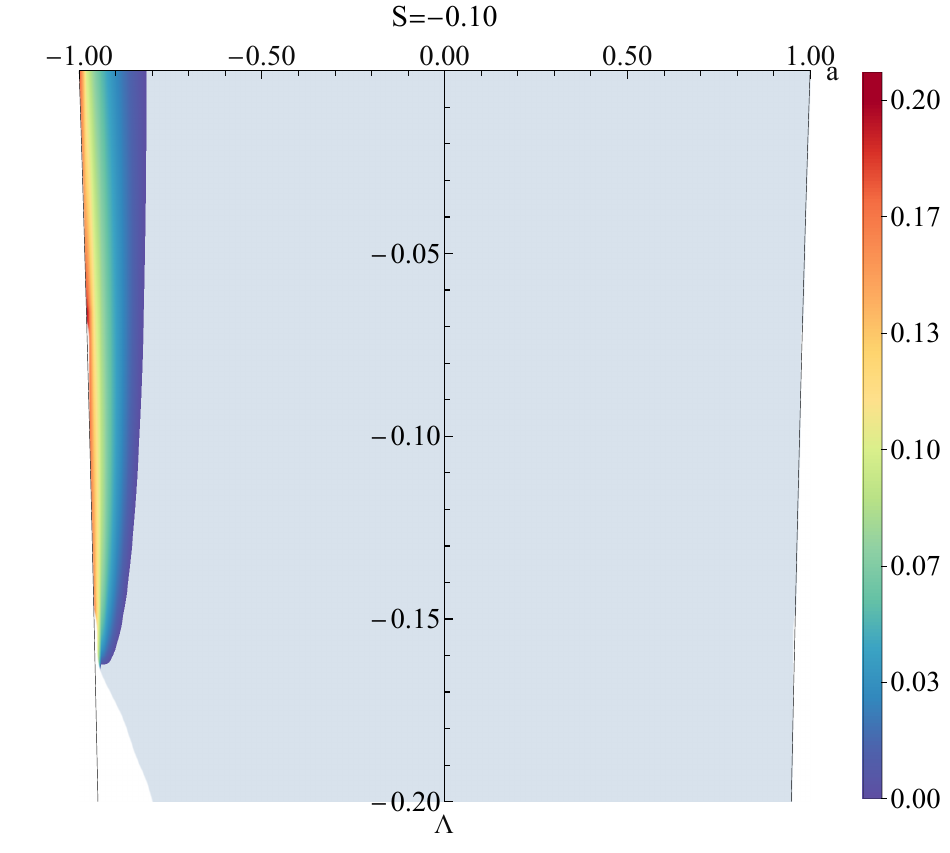}}
		\caption{Dependence of the difference between the LE and surface gravity on the BH rotation parameter and the cosmological constant for $J=25.00$.}
	\label{Fig6}
\end{figure*}

\begin{figure*}[htbp]
	\centering
	\subcaptionbox{}{\includegraphics[height=0.32\textwidth,width=0.4\textwidth]{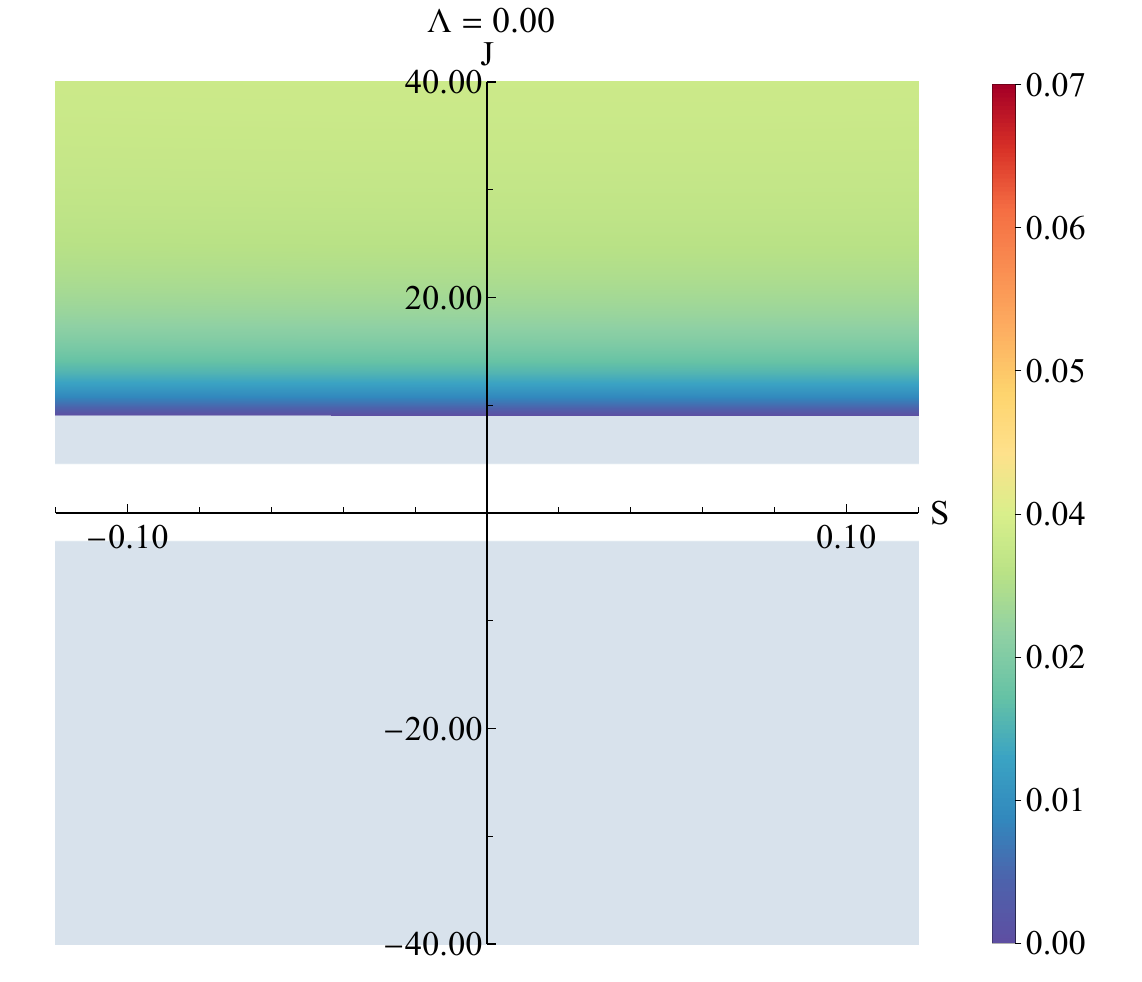}}
	\subcaptionbox{}{\includegraphics[height=0.32\textwidth,width=0.4\textwidth]{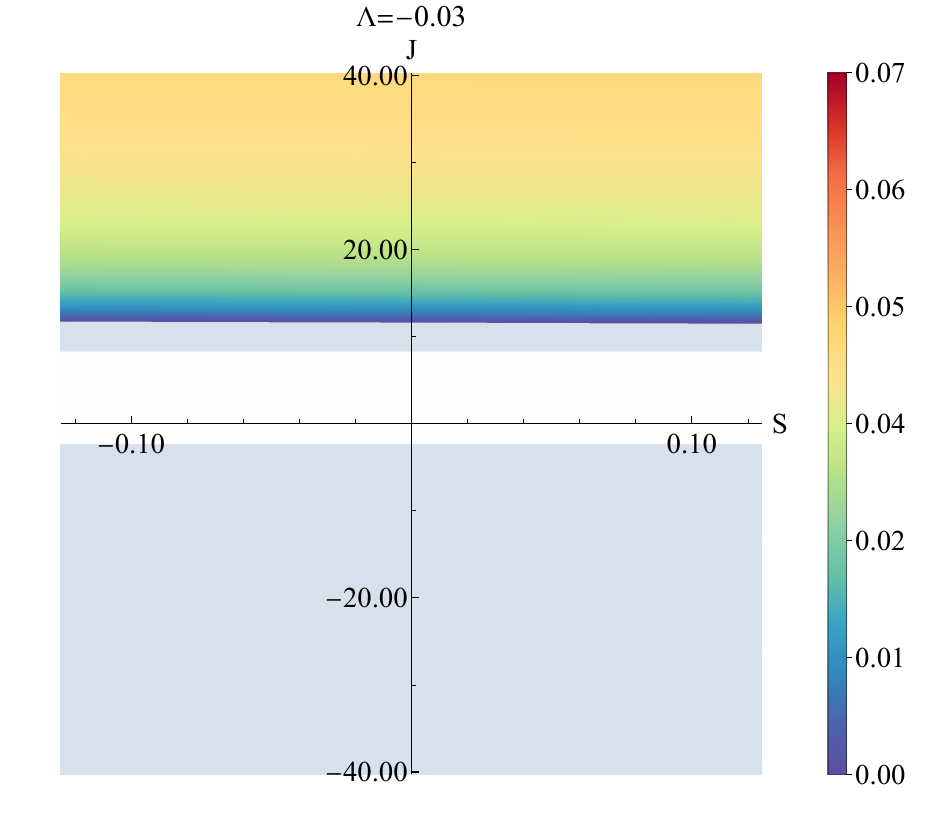}}
	\subcaptionbox{}{\includegraphics[height=0.32\textwidth,width=0.4\textwidth]{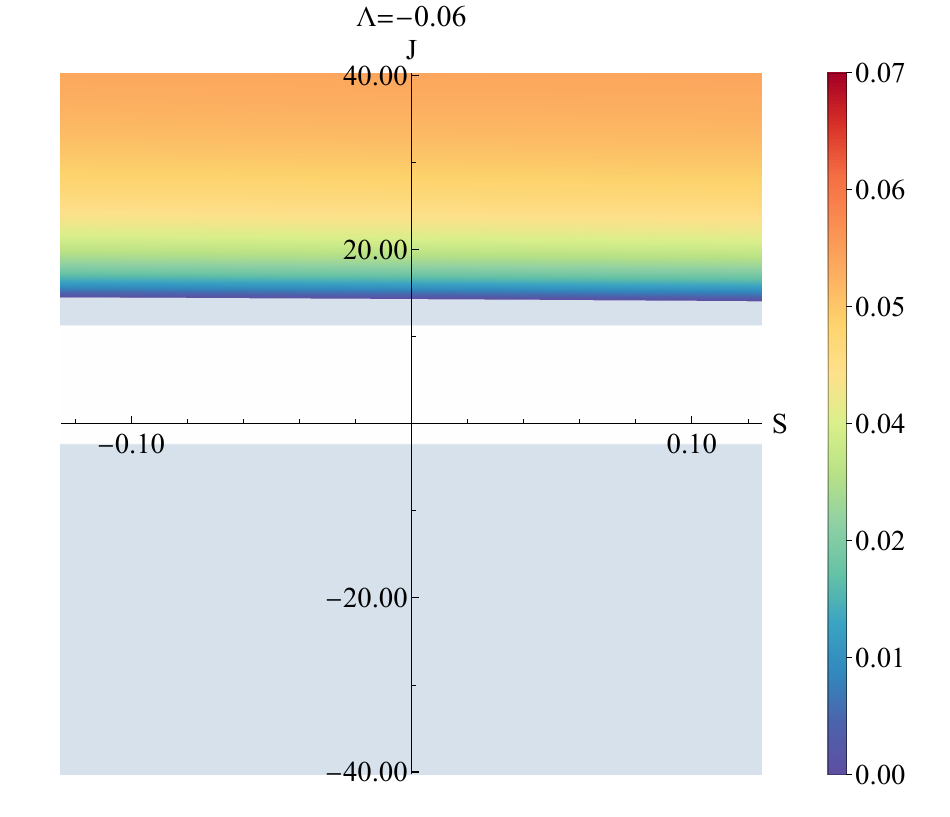}}
	\subcaptionbox{}{\includegraphics[height=0.32\textwidth,width=0.4\textwidth]{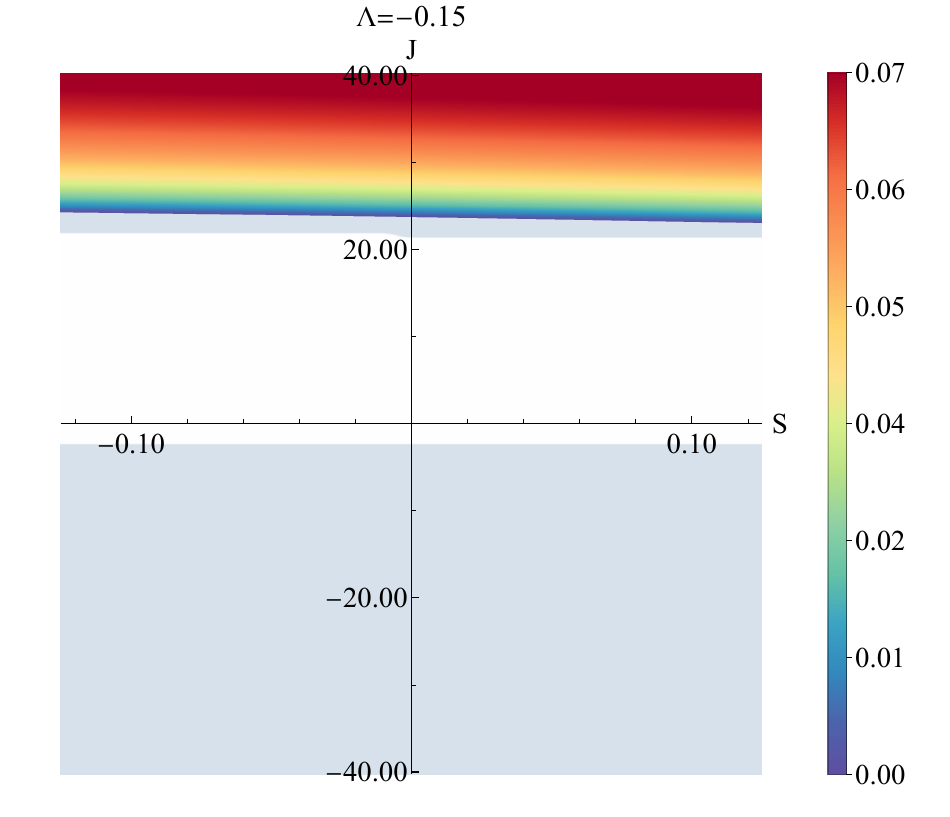}}
	\caption{Dependence of the difference between the LE and surface gravity on the particle spin and total angular momentum for $a=-0.90$.}
	\label{Fig7}
\end{figure*}

\begin{figure*}[htbp]
	\centering
	\subcaptionbox{}{\includegraphics[width=0.32\textwidth]{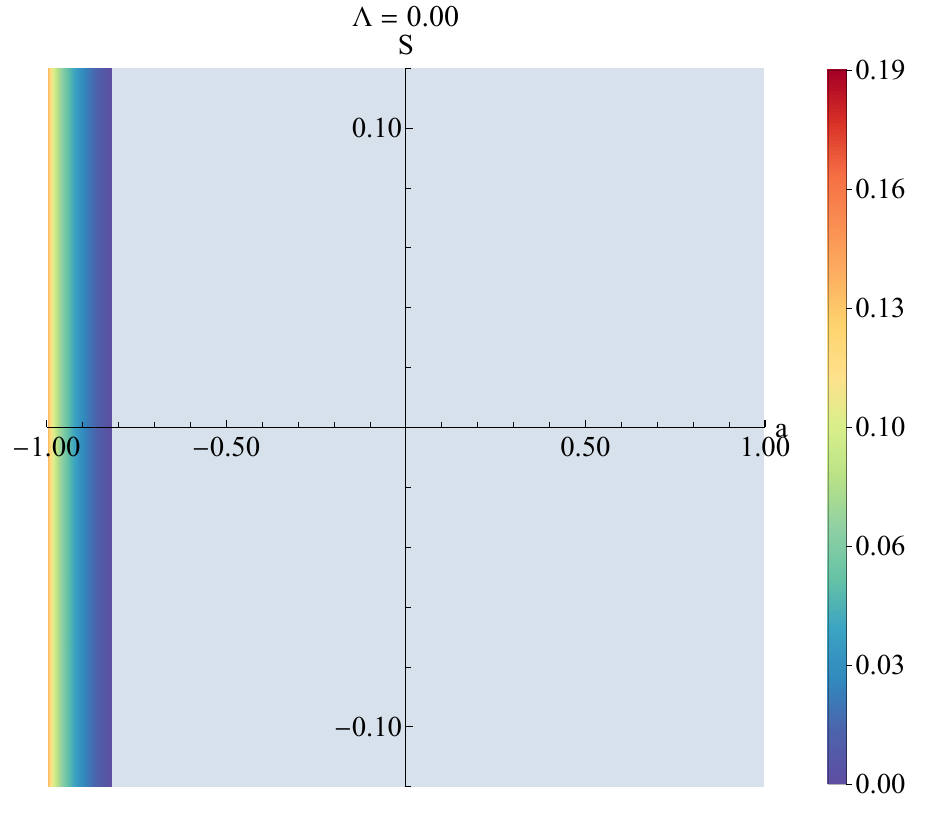}}
	\subcaptionbox{}{\includegraphics[width=0.32\textwidth]{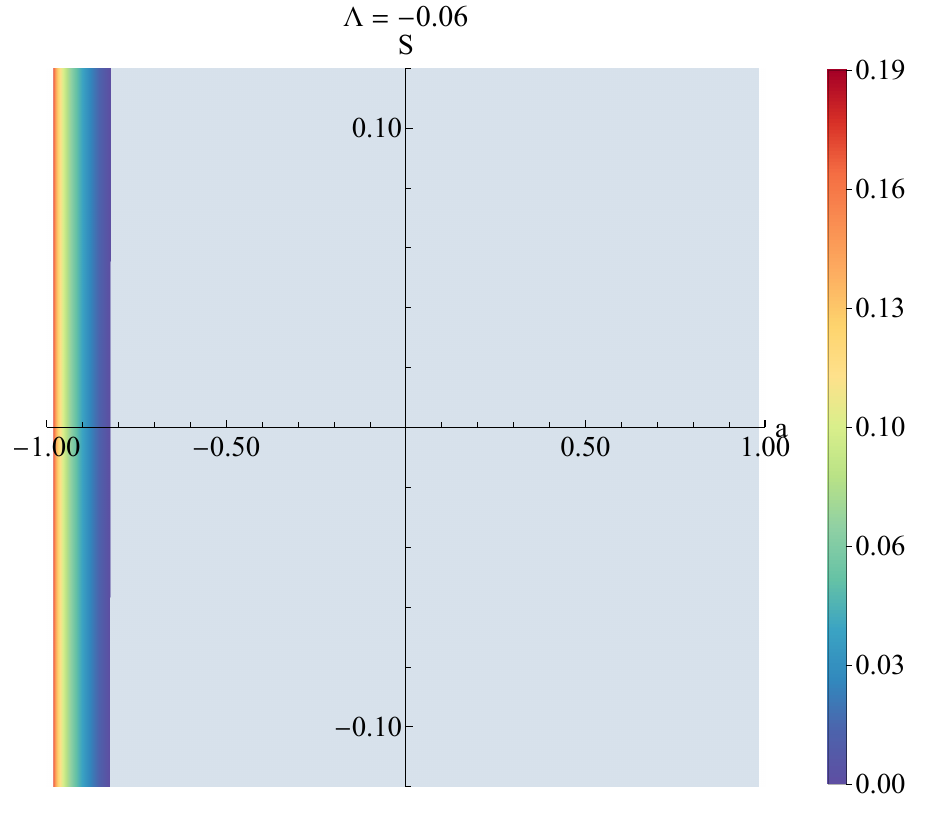}}
	\subcaptionbox{}{\includegraphics[width=0.32\textwidth]{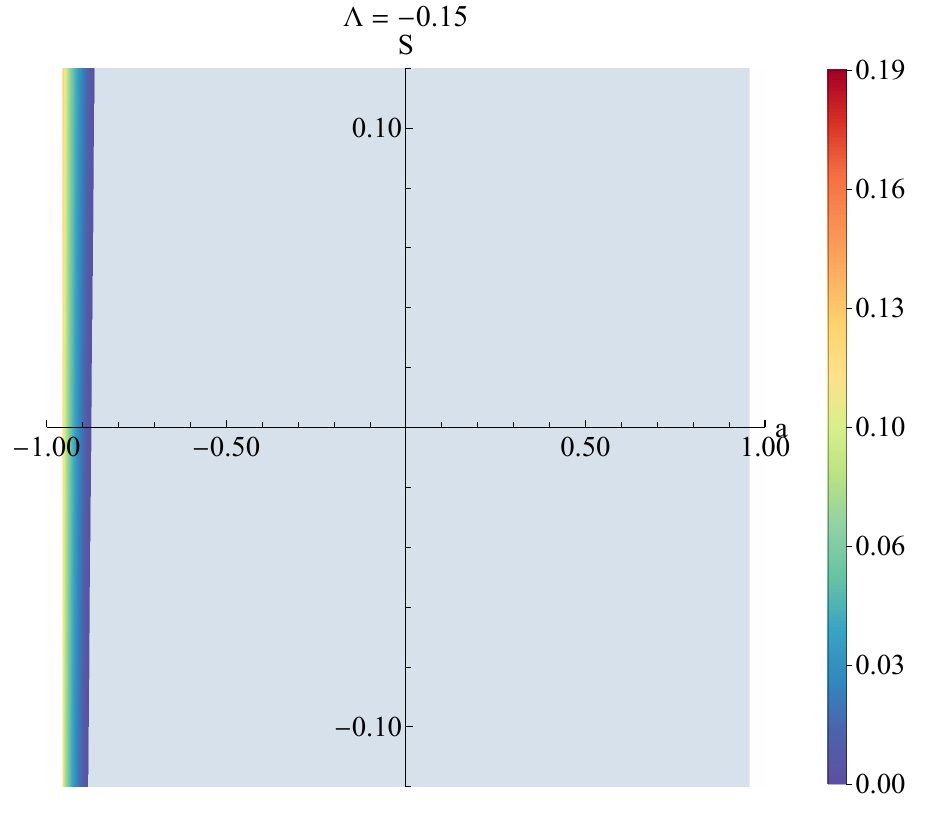}}
	\caption{Dependence of the difference between the LE and surface gravity on the particle spin and the BH rotation parameter for $J=25.00$.}
	\label{Fig8}
\end{figure*}

When the BH charge is set to zero, the Kerr--Newman--AdS spacetime reduces to the Kerr--AdS spacetime. Using Eqs.~\eqref{eq3.2.3} and~\eqref{eq2.3}, we numerically evaluate the LE and the surface gravity in this background. The corresponding results are presented in Figs.~\ref{Fig6}--\ref{Fig8}.

Figure~\ref{Fig6} illustrates the combined effects of the Kerr--AdS BH rotation parameter and the cosmological constant on chaos bound violation for two configurations, $S=0.10$ and $S=-0.10$. In panel~(a), where the particle spin is oriented along the positive $z$-axis, the chaos bound remains satisfied throughout the parameter range with the positive rotation parameter. By contrast, for the rotation in the opposite direction, violation occurs when the magnitude of the rotation parameter is sufficiently large and the magnitude of the negative cosmological constant is small. As the cosmological constant becomes more negative, the violating region progressively shrinks and eventually disappears, reflecting the competition between the local orbital-instability scale and the surface gravity by the AdS background. Panel~(b), corresponding to the case of $S=-0.10$, exhibits the same qualitative dependence on the rotation and cosmological constant, although the parameter region in which the bound is violated is slightly reduced. Thus, the violation shows a pronounced dependence on the direction of BH rotation, while reversing the spin orientation mainly shifts the violation boundary without altering its overall structure.

Figure~\ref{Fig7} illustrates the combined effects of the particle total angular momentum and spin on chaos bound violation for four values of the cosmological constant. Here the BH rotation parameter is negative. When the particle total angular momentum is directed opposite to the positive $z$-axis and is therefore aligned with the BH rotation, the bound remains satisfied throughout the parameter range considered. By contrast, when the total angular momentum is directed along the positive $z$-axis, corresponding to the anti-aligned configuration, the bound is satisfied only at sufficiently small angular momentum magnitudes, while violation occurs once $|J|$ exceeds a critical value. As the cosmological constant becomes more negative, the critical magnitude $|J_c|$ on the anti-aligned branch shifts to larger values, indicating that a larger particle angular momentum is required for the onset of violation. Increasing the particle spin in the positive direction produces the opposite tendency, reducing $|J_c|$ and thereby facilitating violation. These features reflect the asymmetric response of the two angular momentum branches to the rotating background. The spin further modifies the threshold through spin--curvature coupling, while the AdS geometry changes both the local instability of the unstable circular orbit and the BH surface gravity. The resulting violation boundary is therefore determined by the relative response of these two scales rather than by a uniform enhancement or suppression of the tendency toward chaos bound violation.

Figure~\ref{Fig8} illustrates the combined effects of the particle spin and BH rotation on the  violation. In panel~(a), with $\Lambda=0$, violation occurs only for negative values of the rotation parameter, corresponding to the BH rotation opposite to the positive $z$-direction. Within the relatively narrow spin range considered here, variations in either the magnitude or the orientation of the particle spin produce only minor shifts in the boundary separating the bound-satisfying and bound-violating regions. The dependence on the rotation parameter is therefore more pronounced in this parameter regime. As the cosmological constant becomes more negative, the range of rotation parameter associated with the bound violation becomes progressively restricted, as seen in panels~(b) and~(c). This tendency is particularly clear in panel~(c), where no violation is found for $a>-0.88$ within the parameter range considered. Thus, as the cosmological constant becomes more negative, the parameter region in which the LE exceeds the surface gravity scale decreases monotonically, indicating an overall suppression of chaos bound violation.

These results show that the chaos bound behavior arises from the interplay between the background geometry and the dynamics of the spinning particle. The rotation and cosmological constant modify both the effective radial dynamics and the surface gravity, thereby changing the relative balance between the local orbital-instability and surface gravity scales and, consequently, the behavior of the chaos bound. The particle spin and total angular momentum provide additional contributions through spin--curvature coupling and rotational effects. Consequently, the onset of chaos bound violation is governed by the combined influence of the BH background and the dynamical parameters of the particle rather than by any single parameter alone.

\subsection{RN–AdS spacetime}\label{sec3.4}

\begin{figure*}[htbp]
	\centering
	\subcaptionbox{}{\includegraphics[width=0.4\textwidth]{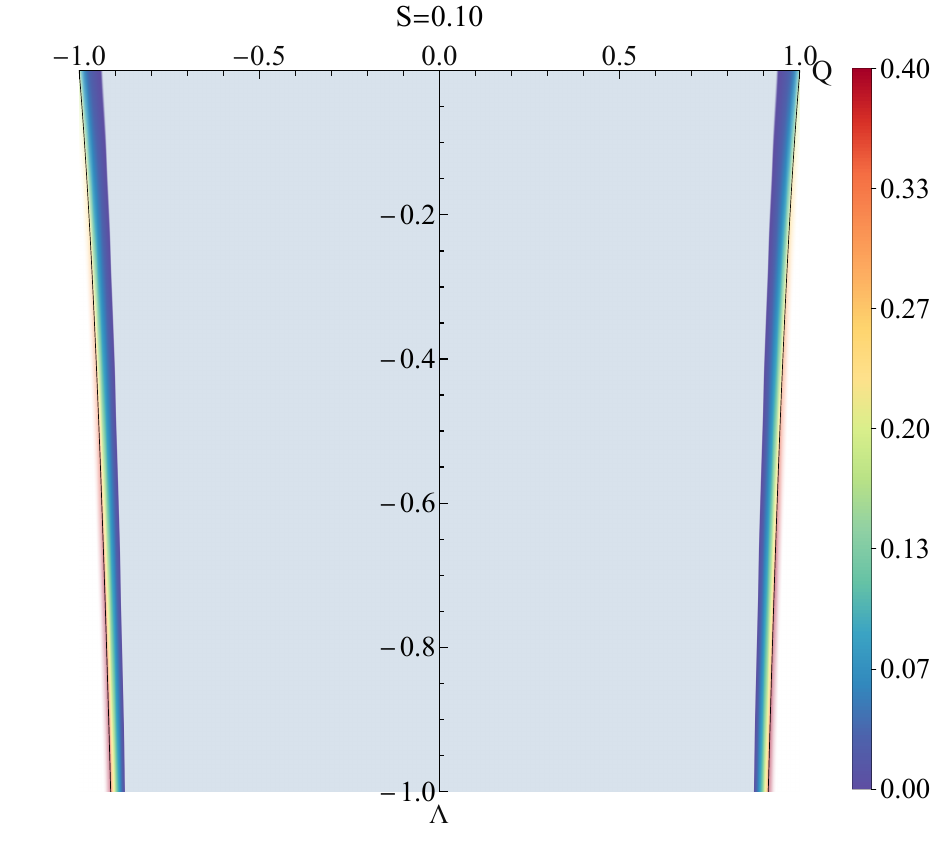}}
	\subcaptionbox{}{\includegraphics[width=0.4\textwidth]{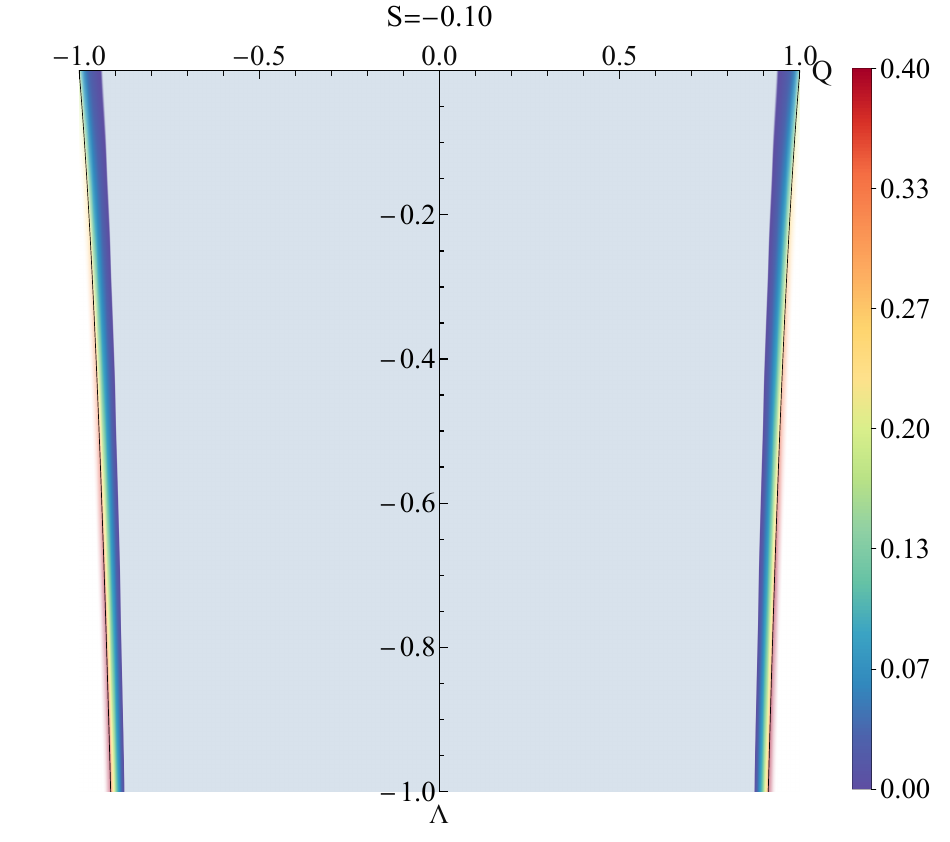}}
	\caption{Dependence of the difference between the LE and surface gravity on the BH charge and the cosmological constant for $J=25.00$ and $q=0.10$.}
	\label{Fig9}
\end{figure*}

\begin{figure*}[htbp]
	\centering
	\subcaptionbox{}{\includegraphics[height=0.32\textwidth,width=0.4\textwidth]{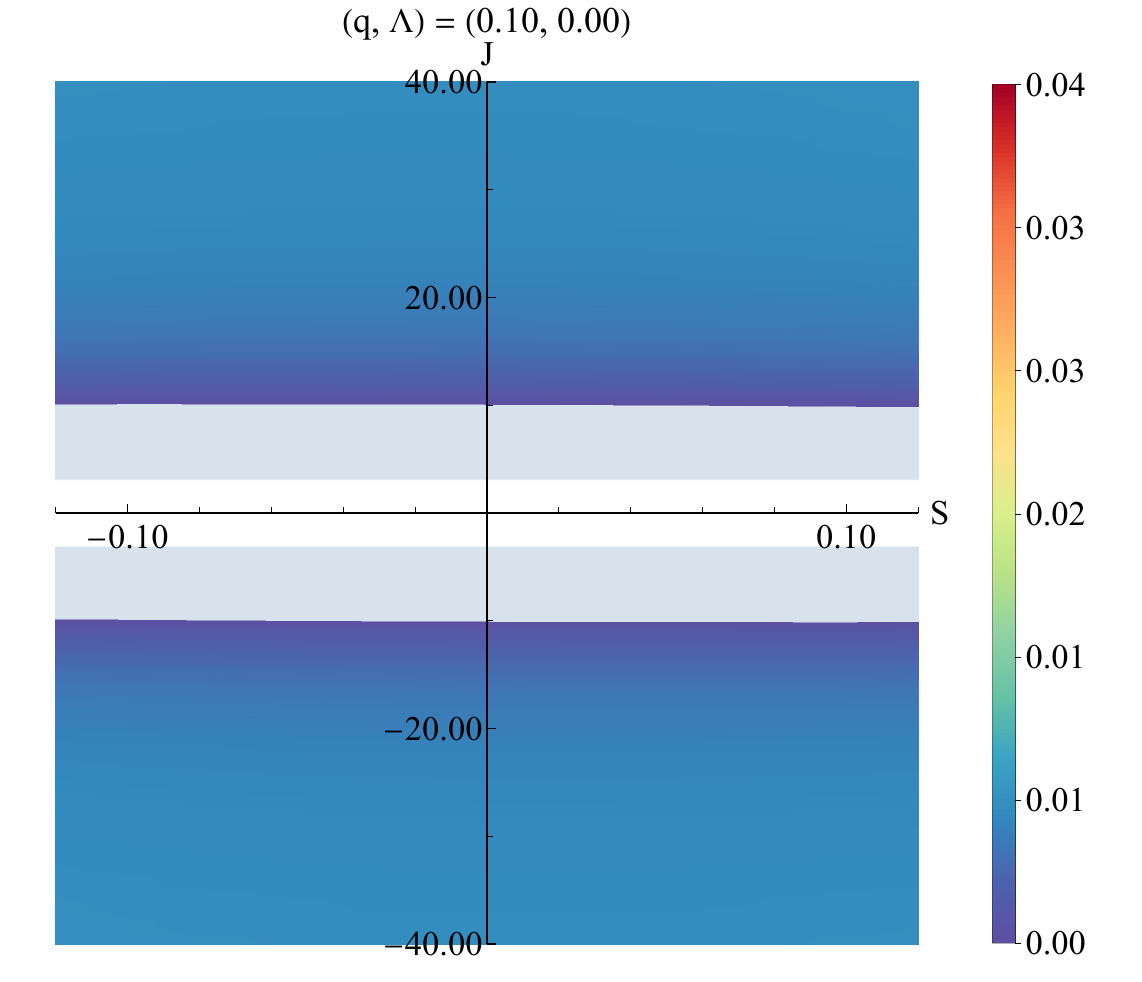}}	\subcaptionbox{}{\includegraphics[height=0.32\textwidth,width=0.4\textwidth]{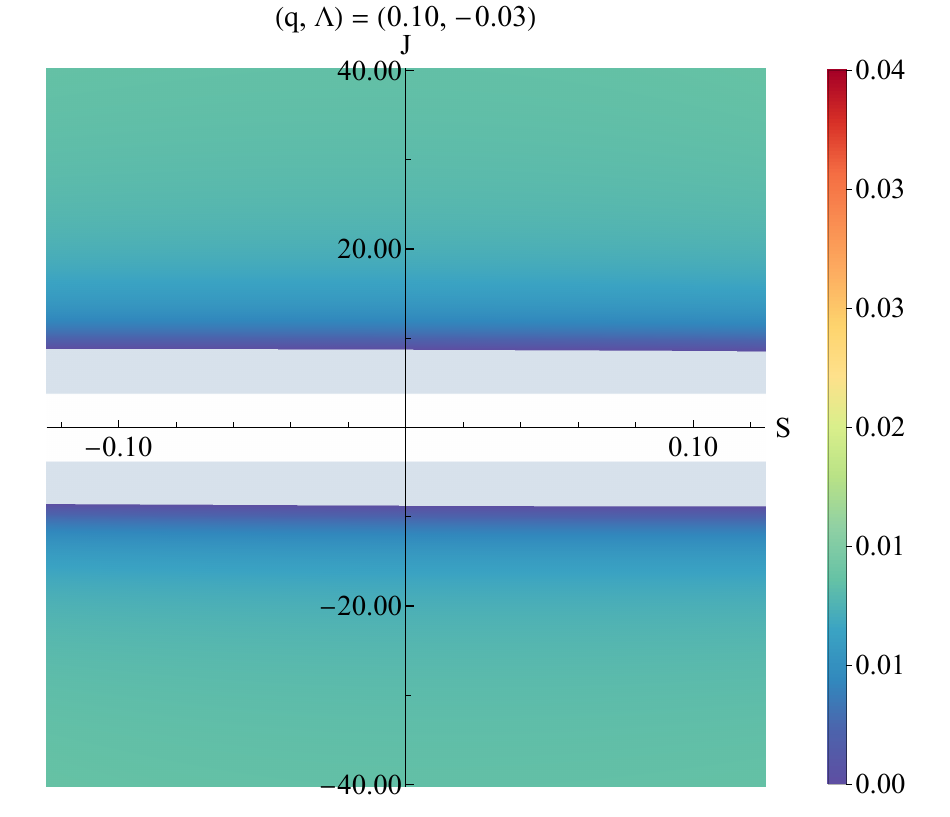}}
	\subcaptionbox{}{\includegraphics[height=0.32\textwidth,width=0.4\textwidth]{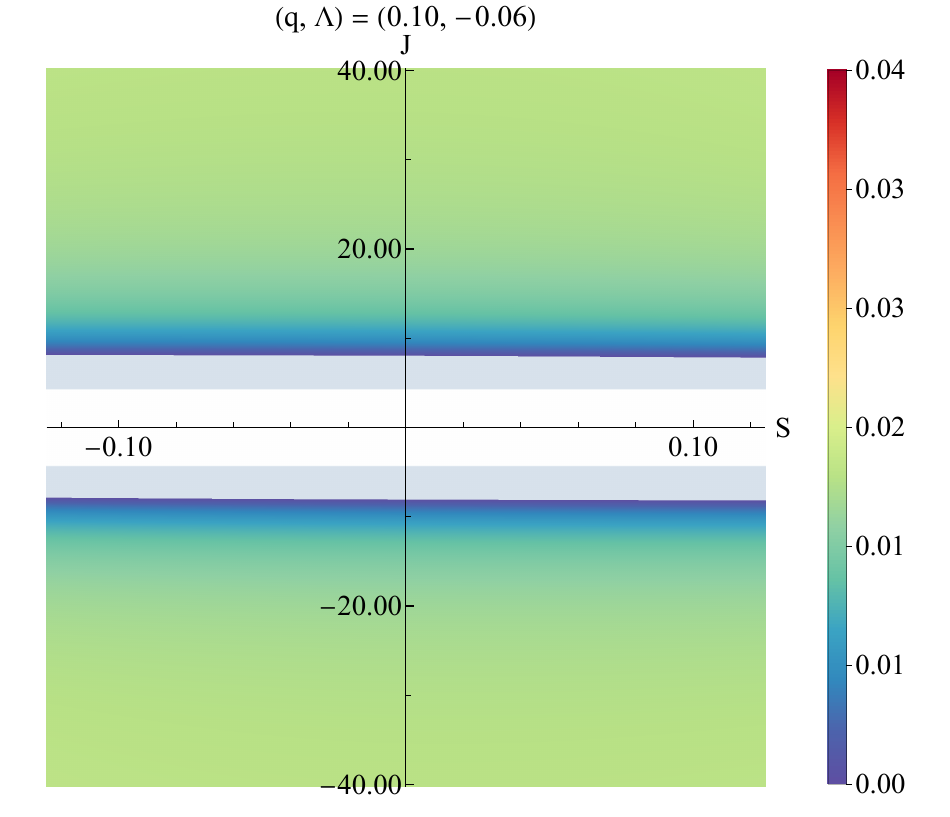}}
	\subcaptionbox{}{\includegraphics[height=0.32\textwidth,width=0.4\textwidth]{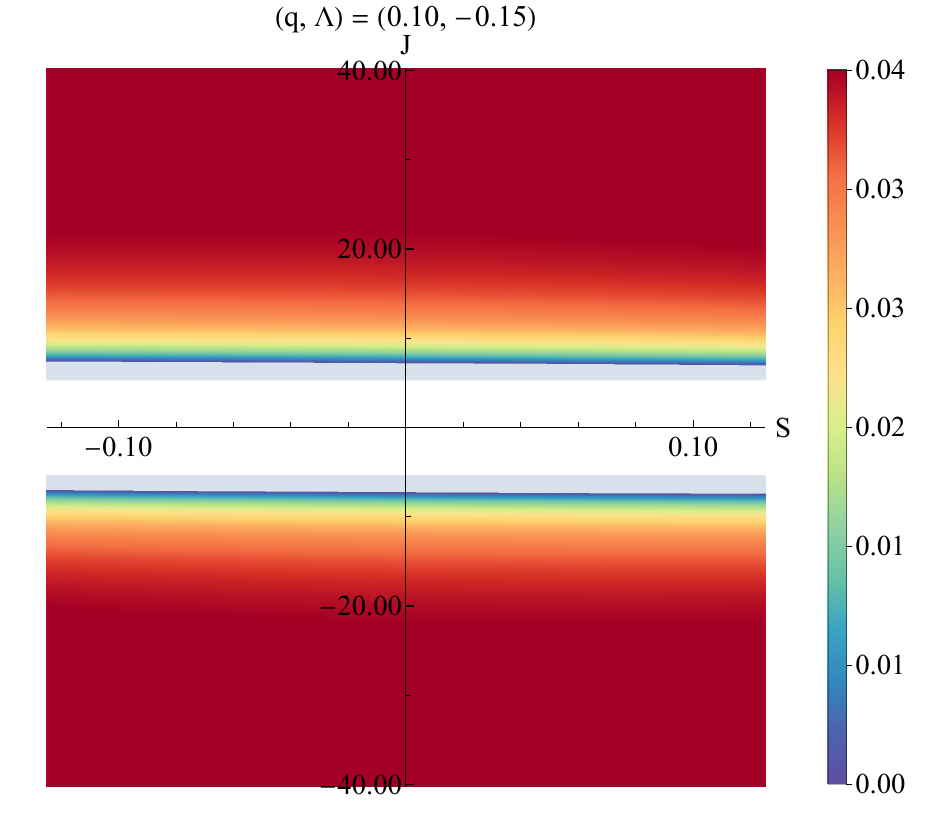}}\\[5mm]
	\subcaptionbox{}{\includegraphics[height=0.32\textwidth,width=0.4\textwidth]{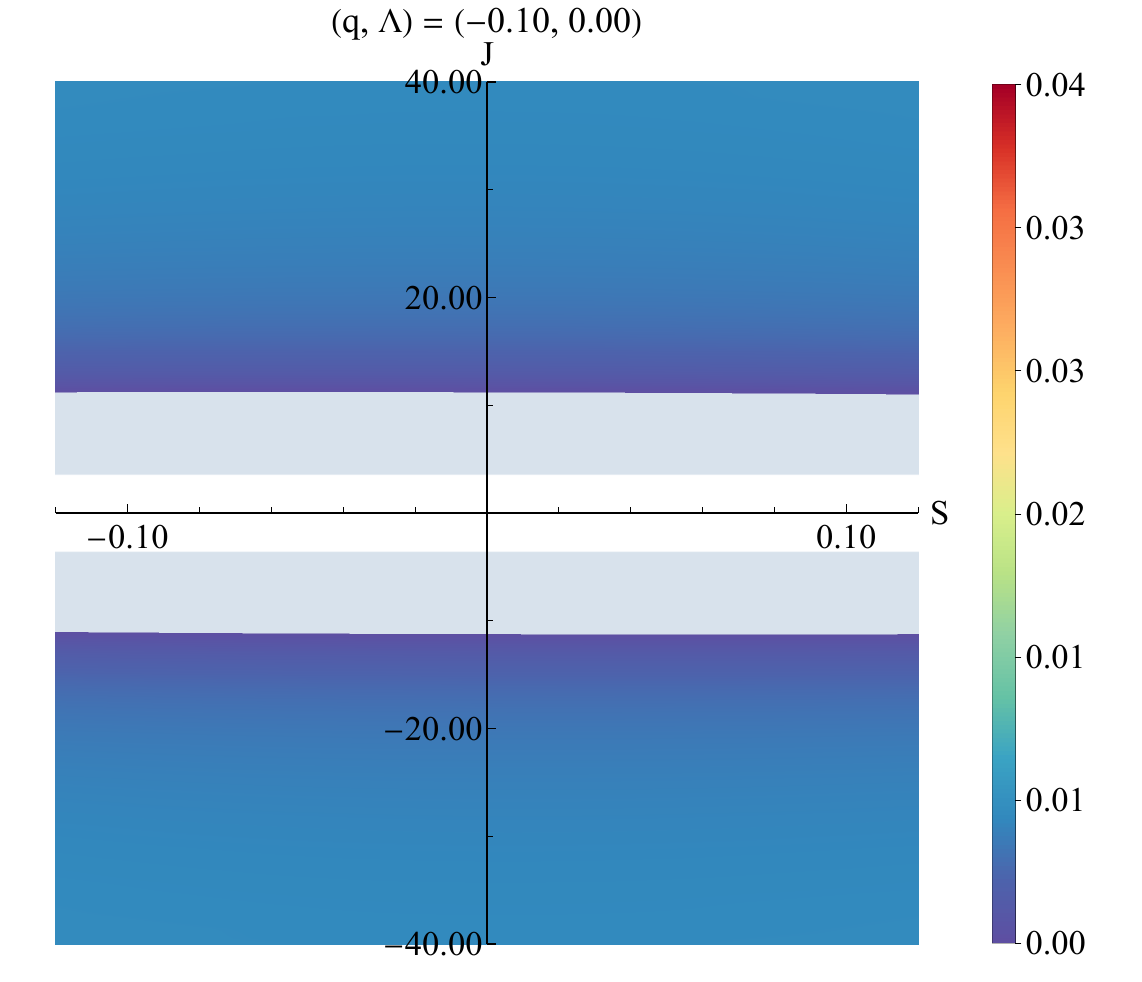}}
	\subcaptionbox{}{\includegraphics[height=0.32\textwidth,width=0.4\textwidth]{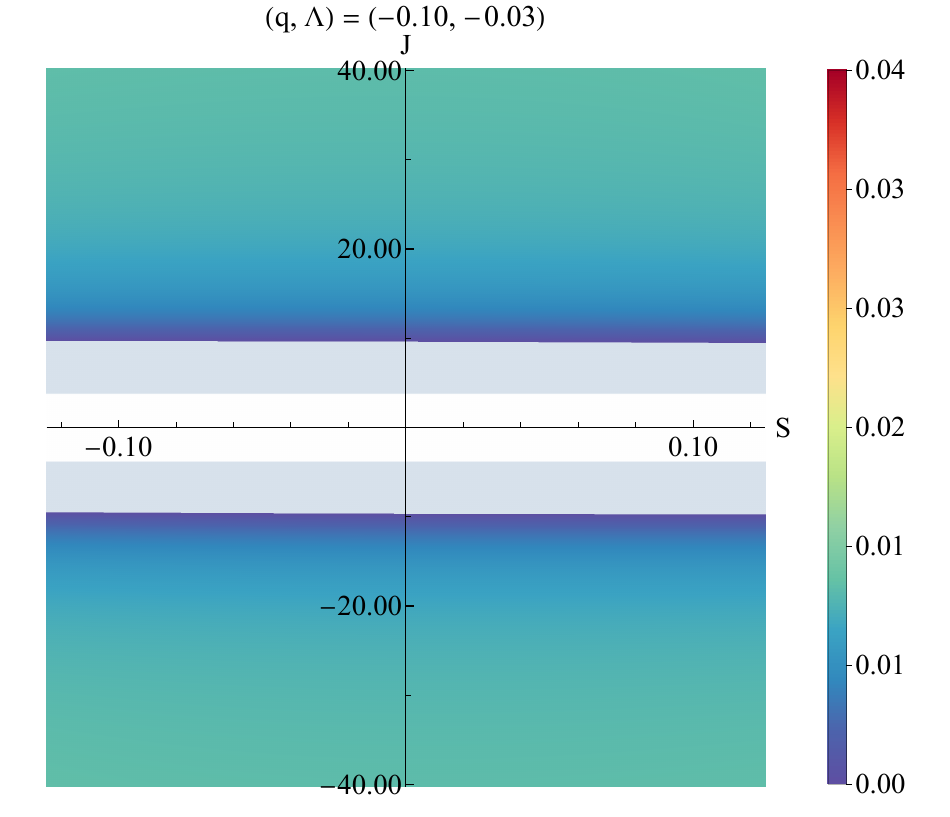}}
	\subcaptionbox{}{\includegraphics[height=0.32\textwidth,width=0.4\textwidth]{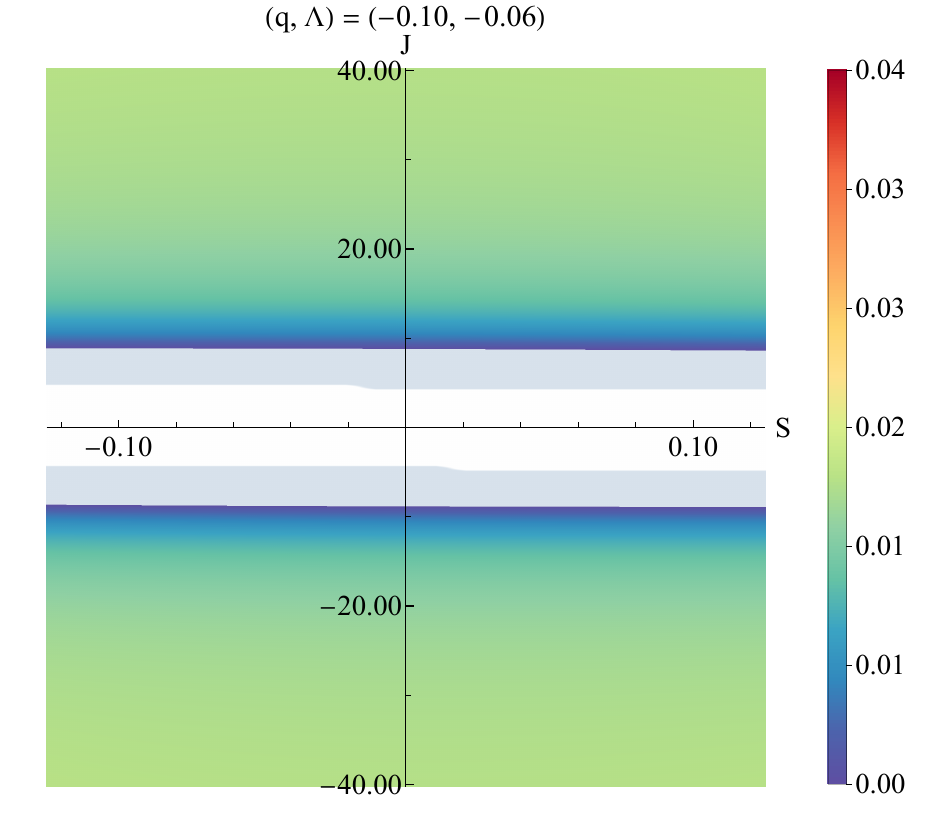}}
	\subcaptionbox{}{\includegraphics[height=0.32\textwidth,width=0.4\textwidth]{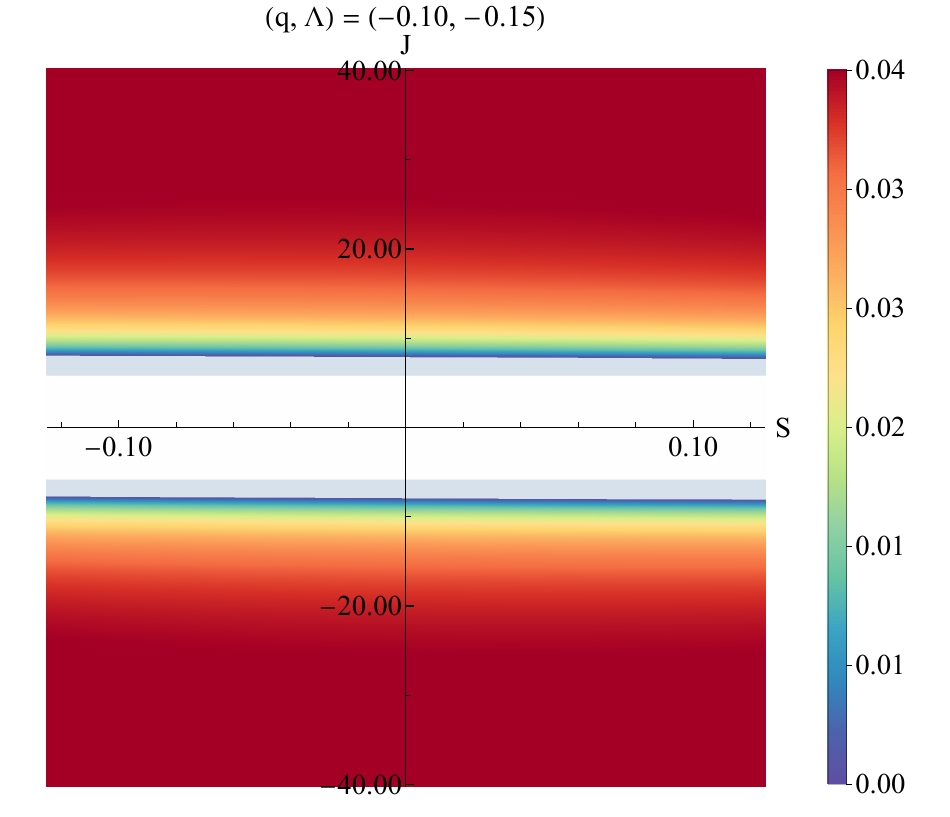}}
	\caption{Dependence of the difference between the LE and surface gravity on the particle spin and total angular momentum for $Q=0.95$.}
	\label{Fig10}
\end{figure*}

When the rotation parameter is set to zero, the Kerr--Newman--AdS spacetime reduces to the RN--AdS geometry. Using Eqs.~\eqref{eq3.2.3} and \eqref{eq2.3}, we then evaluate the LE and surface gravity numerically in this nonrotating limit and examine whether the chaos bound is satisfied. The corresponding results are presented in Figs.~\ref{Fig9} and \ref{Fig10}.

Figure~\ref{Fig9} shows the combined effects of the BH charge and cosmological constant on the chaos bound violation in the nonrotating RN--AdS background. Panel~(a) corresponds to a particle spin of $S=0.10$. As the cosmological constant becomes more negative, the critical charge for the onset of violation decreases, while the corresponding violating interval in charge becomes progressively narrower. Within the parameter range considered, the relatively small particle charge results in a weaker electromagnetic influence on the orbital dynamics than that associated with the particle angular momentum. Nevertheless, a slight asymmetry persists between configurations with opposite charge signs, reflecting the distinct effects of attractive and repulsive electromagnetic interactions. Panel~(b) shows the corresponding result for $S=-0.10$. Compared with panel~(a), the overall structure of the parameter space changes only slightly, indicating that reversing the particle spin within the range considered has only a modest effect on both the onset and extent of chaos bound violation.

Figure~\ref{Fig10} illustrates the combined effects of the particle spin and total angular momentum on the violation in the RN--AdS spacetime. Panel~(a) corresponds to the asymptotically flat RN limit, $\Lambda=0$. For a spinless particle, $S=0$, the total angular momentum coincides with the orbital angular momentum. Owing to the spherical symmetry of the background, reversing its direction, $J\rightarrow -J$, leaves the LE unchanged. For nonzero spin, the spin--curvature coupling introduces a small difference between the two angular momentum branches for a fixed spin orientation, although they remain nearly symmetric over the spin range considered here. This shows that the spin induced correction to the orbital-instability is comparatively weak. As the cosmological constant becomes more negative, as shown in panels~(b)--(d), the chaos bound boundary defined by $\lambda=\kappa$ shifts toward smaller angular momentum magnitudes on both the positive- and negative-$J$ branches. In other words, the critical magnitudes $|J_c^{+}|$ and $|J_c^{-}|$ both decrease as the magnitude of the negative cosmological constant increases, so that the violation sets in at smaller values of $|J|$. This trend indicates that, for the fixed background parameters considered here, a more negative cosmological constant lowers the angular momentum threshold rat which the LE exceeds the surface gravity. The nearly symmetric response of the two branches is consistent with the absence of BH rotation and hence of any preferred azimuthal direction. The displacement of the boundary nevertheless reflects the combined response of the local orbital-instability and the surface gravity to the AdS curvature, rather than a change in either quantity alone.

Panels~(e)--(h) show the corresponding results after reversing the sign of the particle charge, thereby changing the electromagnetic interaction from repulsive to attractive. Relative to the repulsive case, the onset of chaos bound violation is generally shifted toward larger angular momentum magnitudes over the parameter range considered, indicating that the attractive interaction delays the onset of violation. The dependence on the particle spin remains weak throughout all eight panels. Thus, within the parameter range explored here, the violation boundary is more sensitive to the particle total angular momentum, the cosmological constant, and the sign of the electromagnetic interaction, while the particle spin produces only a comparatively small correction. This trend is specific to the nonrotating RN--AdS background and should not be interpreted as a universal effect of the cosmological constant in rotating spacetimes.

\section{Conclusion and discussion}\label{sec4}

In this work, we investigated the chaos bound behavior of spinning charged test particles in the Kerr--Newman--AdS spacetime by comparing the radial LE of unstable circular orbits with the BH surface gravity. Our results show that the occurrence and parameter dependence of the violation reflect the interplay between the background geometry and particle dynamics.

In the Kerr--Newman--AdS spacetime, the violation depends sensitively on the relative orientation between the particle total angular momentum and the BH rotation. Within the parameter range considered, anti-aligned configurations generally favor violation, whereas the bound is more readily satisfied for aligned configurations. The BH charge significantly modifies the violation region but is not a necessary condition for violation, since the bound can also be violated in the Kerr--AdS limit. The particle spin shifts the relevant boundaries through spin--curvature coupling, although its influence remains comparatively weak over the spin range explored here. The effect of the cosmological constant is likewise background and branch dependent, reflecting its simultaneous influence on the local orbital-instability and surface gravity scales. This dependence becomes particularly clear in the limiting geometries. In the Kerr--AdS spacetime, violation is favored for sufficiently strong rotation opposite to the particle total angular momentum, while the corresponding violation region becomes more restricted as the cosmological constant becomes more negative. In the RN--AdS spacetime, where the BH rotation vanishes and frame dragging is absent, the positive- and negative-$J$ branches become nearly symmetric. As the cosmological constant becomes more negative, the critical magnitude of the particle total angular momentum required for violation decreases on both branches. These contrasting trends indicate that the cosmological constant does not produce a universal enhancement or suppression of chaos bound violation, but instead modifies the relevant parameter boundaries in a manner that depends on the background geometry and the parameter branch considered.

Our analysis was carried out within the test particle approximation and over a relatively narrow range of particle spin. The LE considered here characterizes radial instability within the symmetry-reduced equatorial sector and does not address the stability of the aligned or anti-aligned spin configurations under generic spin-tilt or off-equatorial perturbations. Addressing such perturbations would require a coupled analysis of the full MPD dynamics. Backreaction and possible semiclassical or quantum corrections are likewise beyond the scope of the present work and provide natural directions for future investigation.

\end{document}